\documentclass[a4paper,12pt]{article}
\usepackage[margin=0.6in]{geometry}

\usepackage{graphicx}
\usepackage{epstopdf}
\usepackage{amssymb,amsmath}
\usepackage[colon,authoryear]{natbib}
\pdfoutput=1
\begin{document}

\title{Inclined asymmetric librations in exterior resonances}

\author{G. Voyatzis$^1$, K. Tsiganis$^1$, K. I. Antoniadou$^2$\vspace{0.25cm}\\
\small{$^1$Department of Physics, Aristotle University of Thessaloniki, 54124,
              Thessaloniki, Greece} \vspace{-0.15cm}\\
							\small{$^2$NaXys, Department of Mathematics, University of Namur, 8 Rempart de la Vierge, 5000 Namur, Belgium}\vspace{0.25cm}\\
              \small{voyatzis@auth.gr, tsiganis@auth.gr, kyriaki.antoniadou@unamur.be} }    
		
		\date{}
		\maketitle
		
\begin{center}
The final publication is available at\\ https://link.springer.com/article/10.1007/s10569-018-9821-0
\end{center}
\begin{abstract}
Librational motion in celestial mechanics is generally associated with the existence of stable resonant configurations and signified by the existence of stable periodic solutions and oscillation of critical (resonant) angles.  When such an oscillation takes place around a value different than 0 or $\pi$, the libration is called {\it asymmetric}. In the context of the planar circular restricted three-body problem (CRTBP), asymmetric librations have been identified for the exterior mean-motion resonances (MMRs) $1:2$, $1:3$ etc.~ as well as for co-orbital motion ($1:1$). In exterior MMRs the massless body is the outer one. In this paper, we study asymmetric librations in the 3-dimensional space. We employ the computational approach of Markellos (1978) and compute families of asymmetric periodic orbits and their stability. Stable, asymmetric periodic orbits are surrounded in phase space by domains of initial conditions which correspond to stable evolution and librating resonant angles. Our computations were focused on the spatial circular restricted three-body model of the Sun-Neptune-TNO system (TNO~=~ trans-Neptunian object). We compare our results with numerical integrations of observed TNOs, which reveal that some of them perform $1:2$-resonant, inclined asymmetric librations. For the stable $1:2$ TNOs librators, we find that their libration seems to be related with the vertically stable planar asymmetric orbits of our model, rather than the 3-dimensional ones found in the present study.

\end{abstract}
{\bf keywords} circular restricted TBP --  exterior resonances -- spatial asymmetric periodic orbits -- trans-Neptunian object dynamics

\section{Introduction}
In our Solar system,  exterior MMRs are important for the dynamics of small bodies that move exterior to the orbit of a giant planet and especially exterior to Neptune's orbit, where many TNOs exist in the Kuiper belt and beyond in a scattered disk \citep{Jewitt99}. An exceptional dynamical phenomenon, which occurs in exterior resonances of the form $1:p$ (where $p$=2, 3,..), is the existence of asymmetric librations, namely orbits with a resonant argument, $\sigma$, that librates around a value different than  $0$ or $\pi$. Capture of particles and small bodies in such resonances is possible under the effect of drag \citep{bf94} and a number of TNOs seems to be located in an $1:2$ resonant asymmetric configuration \citep{LykawkaMukai07}.    

In the context of the planar circular three-body problem (CRTBP), the centres of such librations in phase space consist of linearly stable asymmetric periodic orbits as it is revealed clearly in Poincar\'e surfaces of section \citep{winter97,vkh05}. The existence of asymmetric periodic orbits (denoted hereinafter by {\em a.p.o.}) has been shown by \cite{Message58} for the $1:2$ resonance between a Jovian-like planet and a small body (a potential asteroid). Extensive numerical studies of a.p.o. in the planar CRTBP have been performed by \cite{Taylor83a,Taylor83b}, who showed the existence of bifurcations of families of a.p.o. for all exterior MMRs with $p$ up to 51 and computed whole families ($p\leq 5$) or particular segments ($p\leq 12$). \cite{Beauge94} used an analytical approach to show the existence of asymmetric librations in $1:p$ exterior MMRs and their absence in other exterior MMRs as e.g. the $2:3$ and $3:4$. We should remark that i) a.p.o are found also in the elliptic planar restricted and in the general planetary planar problem \citep{avk11} ii) $1:1$ families of a.p.o. also exist and emanate from $L_4$ for the circular planar restricted \citep{Zagouras96} and these families continue in the general planetary problem \citep{Giuppone10, hv11}.        

Concerning spatial three-body models, families of symmetric periodic orbits ({\em s.p.o.}) have been mainly computed for asteroids and TNOs \citep{ikm89, Hadjidem93, kv05} or for exoplanetary systems \citep{av13b,av14b}. Generally, families of s.p.o. bifurcate from planar periodic orbits which are critical with respect to their vertical stability \citep{henon73} and are called {\em vertical critical orbits} (v.c.o.). These families extend up to some critical value of the inclination, $i$, or terminate at $i=180^\circ$ (planar retrograde orbit). With respect to their linear stability, the studies cited above showed that most prograde orbits of moderate or high inclination are unstable. Nevertheless, segments of stable orbits also exist providing restricted phase space domains, where resonant angles librate around $0$ or $\pi$. \cite{Markellos78} addressed the problem of computation of spatial a.p.o. based on the conjecture that asymmetric v.c.o. consist also bifurcation points for the generation of families of spatial a.p.o. He studied the exterior MMR $1:2$ by computing the asymmetric v.c.o. of the planar problem for mass parameter $\mu \in [0.001,0.5]$ and provided some samples of spatial a.p.o.   
 
In the present study, we use the approach followed by Markellos, as mentioned above, in order to compute $1:p$ ($p=2,3,4,5$) resonant families of a.p.o.~ in three dimensions, which can be related to the dynamics of TNOs. In the framework of the CRTBP (Sun-Planet-asteroid), we present computations by considering Neptune as the planet, but the results are qualitatively similar when we replace Neptune by Saturn or Jupiter. In Sect. 2 we study the planar families of a.p.o., we compute their vertical stability and locate their v.c.o. In Sect. 3, starting from the v.c.o. and by using continuation we compute the families of spatial a.p.o. and their linear stability. The librations and the long-term evolution of orbits near the spatial a.p.o. are studied numerically in Sect. 4. Also, the stability of such solutions under the effect of all giant planets of our Solar system is examined. In Sect. 5, we consider real TNOs, which show asymmetric librations and study the possible relation between their dynamics and a.p.o. Finally, we summarize our results and conclude in Sect. 6.

\section{Planar families of periodic orbits and vertical stability}
We consider a system {\em Sun -- planet -- massless body} described in a rotating frame $Oxyz$ (of period $2\pi$) by the spatial CRTBP with mass parameter $\mu$ (equal to the ratio of the planetary mass over the total mass which is normalized to unity) and the Sun and planet being fixed on the $Ox$ axis at $x=-\mu$ and $x=1-\mu$, respectively (see \cite{sze}, p.\ 557 and \cite{murraydermott}, p.\ 67). Also, the energy integral is defined as $h=-C_J/2$, where $C_J$ is the Jacobi constant.
The equations of motion of the model obey the symmetry 
\begin{equation}\label{EqSymmetry}
\Sigma: t,x,y,z \rightarrow -t,x,-y,z,
\end{equation}
and, if a periodic orbit is invariant under $\Sigma$ then it is symmetric, otherwise it is asymmetric and $\Sigma$ maps the a.p.o. to a different one (mirror image). The spatial model is consistent with the planar one if we consider initial conditions $z(0)=\dot{z}(0)=0$. In this case, an s.p.o. is given by initial conditions $(x_0,y_0=0, \dot x_0=0,\dot y_0)$, while an a.p.o. has $\dot x_0 \neq 0$. 

\subsection{The structure of $1:p$ planar resonant families}
In the planar CRTBP, we can determine the family $C$ of prograde circular periodic orbits with mean motion ratio $n/n'$, where $n$ and $n'=1$ denote the mean motion of the small body and the planet, respectively. Two families  of resonant periodic orbits (denoted by {\em family} $I$ and {\em family} $II$) bifurcate from circular orbits, where $n/n'$ is almost rational. These families consist of s.p.o. and along them the mean motion ratio remains almost constant and close to its rational value. The topology close to the bifurcation shows some differences that depend on the order of the resonance,  \citep[see for details][]{henon97,Bruno94,Hadjidem93}. Also, considering $\mu=5.15\cdot10^{-5}$ (Neptune's mass) families for first, second and third order resonances are given by \citet{vk05}.  

\begin{figure}
$
\begin{array}{ccc}
\includegraphics[width=0.45\textwidth]{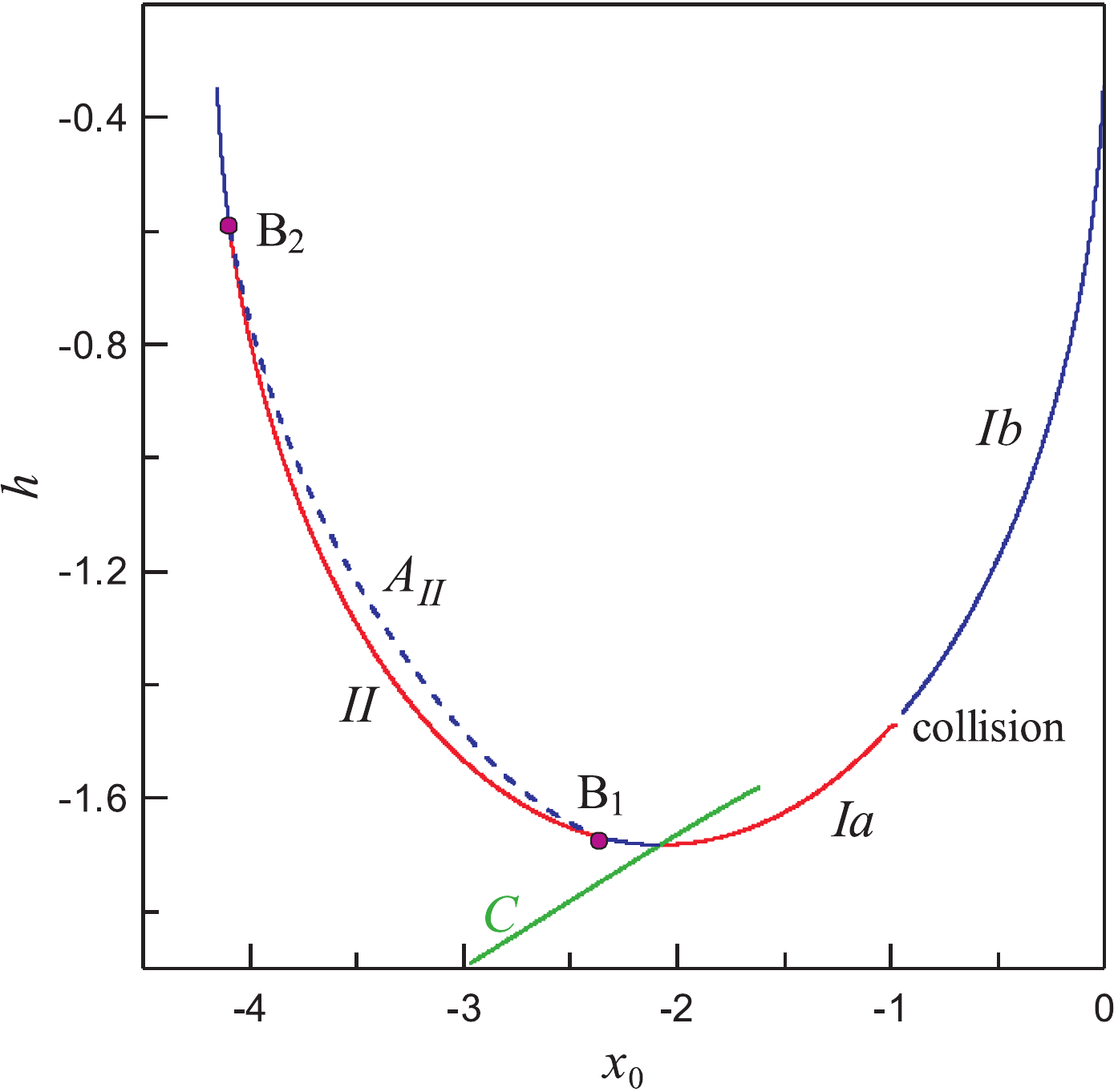} & \quad & \includegraphics[width=0.45\textwidth]{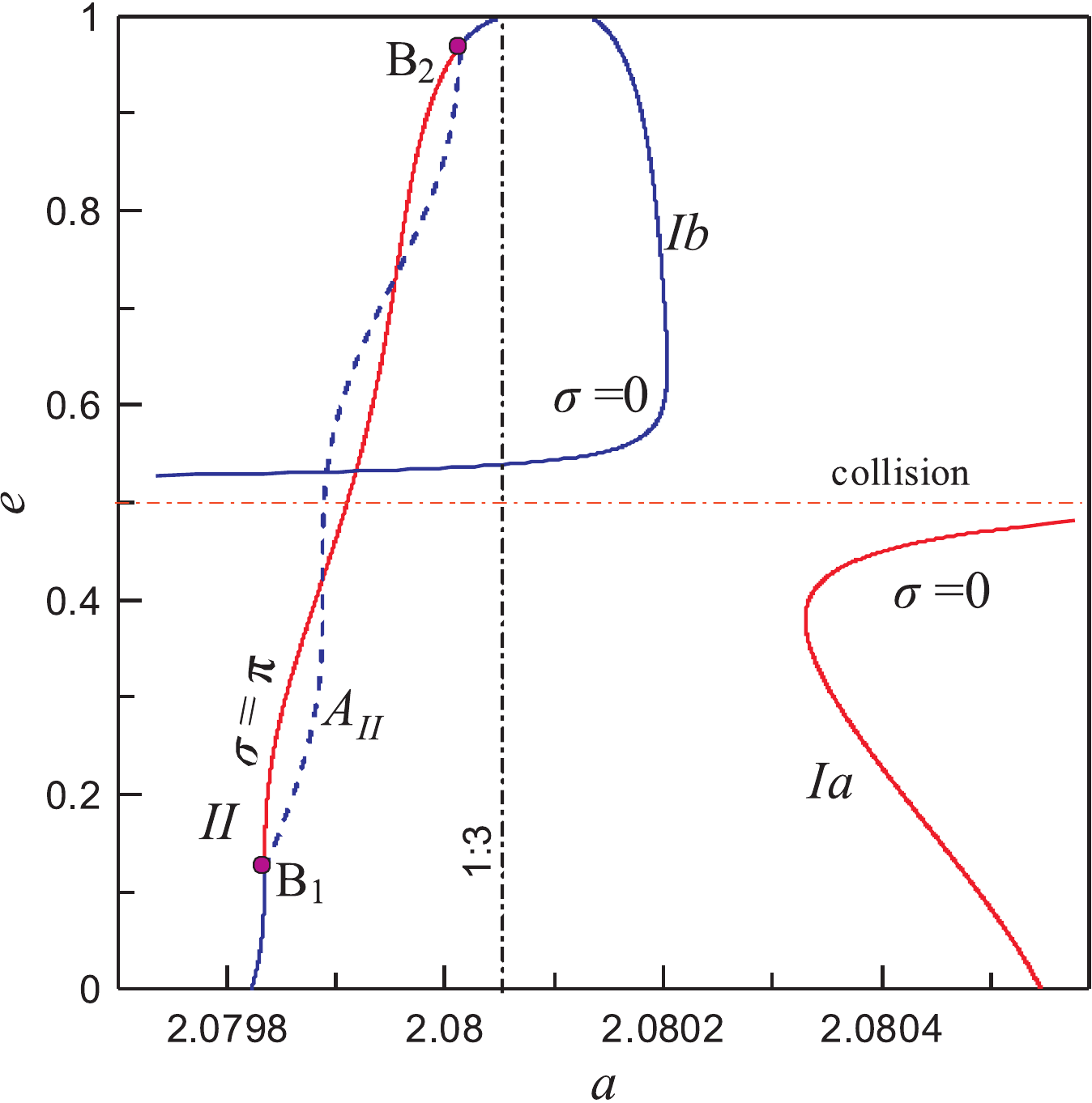} \\
\textnormal{(a)} & \quad & \textnormal{(b)}  
\end{array}$
\caption{Families of planar periodic orbits at the $1:3$ exterior MMR for $\mu=5.15\cdot 10^{-5}$  {\bf a} Families on the $x_0-h$ plane. The characteristic curves of families $I$ and $II$ uniquely determine the initial conditions of the s.p.o., since it is $y_0=\dot x_0=0$, but the asymmetric family $A_{II}$ (dashed line) is a projection, because $\dot x_0\neq 0$. Blue (red) line segments indicate stable (unstable) orbits. The green {\em C} curve is the family of the circular periodic orbits. {\bf b} Families on the $a-e$ plane.  The vertical dot-dashed line indicates the semimajor axis value where the MMR is located in the Keplerian approximation}
\label{FIGPRES13}
\end{figure}

With regard to $1:p$ resonant families, after their bifurcation they are continued towards higher values of both the Jacobi constant, $C_J$ and the eccentricities. As a typical example, we present the families of the $1:3$ exterior MMR for $\mu=5.15\cdot 10^{-5}$ in Fig. \ref{FIGPRES13}. The family branch $Ia$ consists of linearly unstable orbits and terminates at a collision with the planet. For higher energy values, we find the branch $Ib$, which consists of stable orbits and extends up to eccentricity $e=1$ (collision with Sun). Family $II$ starts with stable orbits, but for a critical value of $h$, namely at the orbit $B_1$ (where $e=e_{B1}$), it becomes unstable. At very high eccentricities, in particular at orbit $B_2$ ($e=e_{B2}$), the family becomes again stable and terminates at $e=1$. 

The critical orbits $B_1$ and $B_2$ of the family $II$ are bifurcation points for families of a.p.o. Computations show that the asymmetric family that starts from $B_1$ terminates at $B_2$ and, thus, only one family of planar a.p.o. exists (family $A_{II}$ -- and its mirror image), and it is stable for all values of $h$ \citep{Taylor83a, Bruno94, vkh05}.  Our computations show that this is the case for $\mu<\mu*=5.17\cdot 10^{-3}$. For $\mu=\mu*$ a segment of unstable a.p.o. appears along $A_{II}$ at eccentricity $e*\approx 0.74$ and extends to both lower and higher eccentricities, as $\mu$ increases. The existence of the unstable a.p.o. is followed by a pitchfork bifurcation and the presence of two new asymmetric a.p.o. \cite{Taylor83b} showed the break of this family for $\mu>0.17$ is due to a close encounter with the planet.

\begin{figure}
\centering
\includegraphics[width=0.6\textwidth]{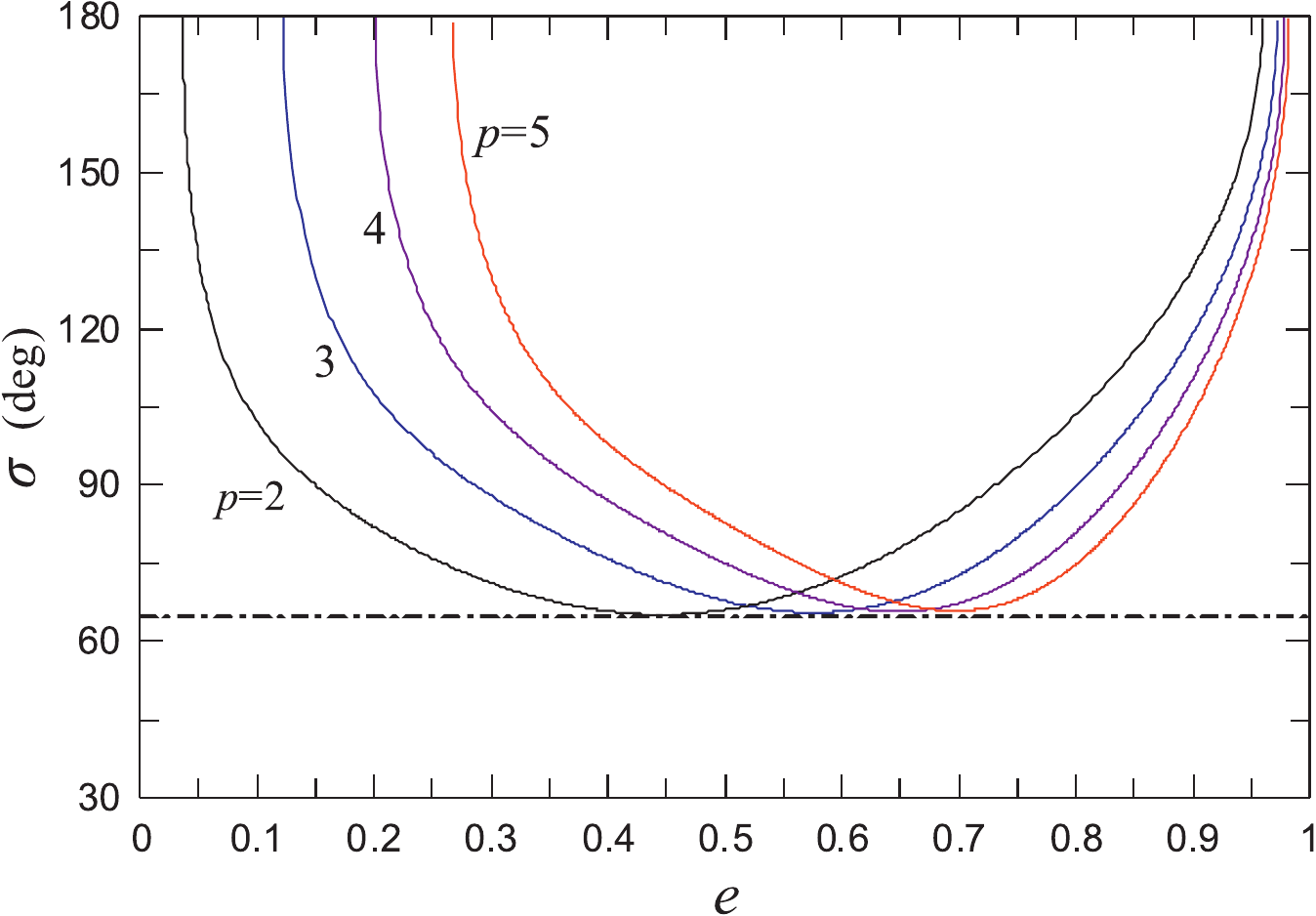}
\caption{The evolution of the resonant argument $\sigma$ along the orbits of the family $A_{II}$. The values $-\sigma$ correspond to the mirror image orbits} 
\label{FIGSIGMA}
\end{figure}  

If we define the resonant angle $\sigma=\lambda'-p\,\lambda+(p-1)\varpi$, where $\lambda$ indicates the mean longitude (primed quantities refer to the planet) and $\varpi$ is the longitude of pericentre, then the s.p.o. of families $I$ and $II$ are located at $\sigma=0$ and $\sigma=\pi$, respectively. However, for the orbits of the asymmetric family $A_{II}$, the resonant angle is not fixed to a given value. The variation of $\sigma$ with $e$ is shown in Fig. \ref{FIGSIGMA}. We note that, for the mirror image orbits, the resonant angle and the pericentre longitude take values $2\pi-\sigma$ and $2\pi-\varpi$, respectively \citep{vkh05}. 

\subsection{Vertical stability of a.p.o.}
Following \cite{henon73}, we consider the vertical variations on planar periodic solutions and define the vertical stability index
$$
b_v=|\textnormal{trace} \Delta(T)|,
$$
where $\Delta(T)$ is the monodromy matrix for the vertical variations computed for one period $T$ of the planar periodic orbit. If $b_v<2$, the variations are bounded and the planar periodic orbit is vertically stable. If $b_v>2$ the planar orbit is vertically unstable and small initial variations from the planar orbit may cause significant variations of the inclination. If $b_v=2$ the orbit is classified as vertical critical orbit (v.c.o.).

\begin{figure}
$
\begin{array}{ccc}
\includegraphics[width=0.45\textwidth]{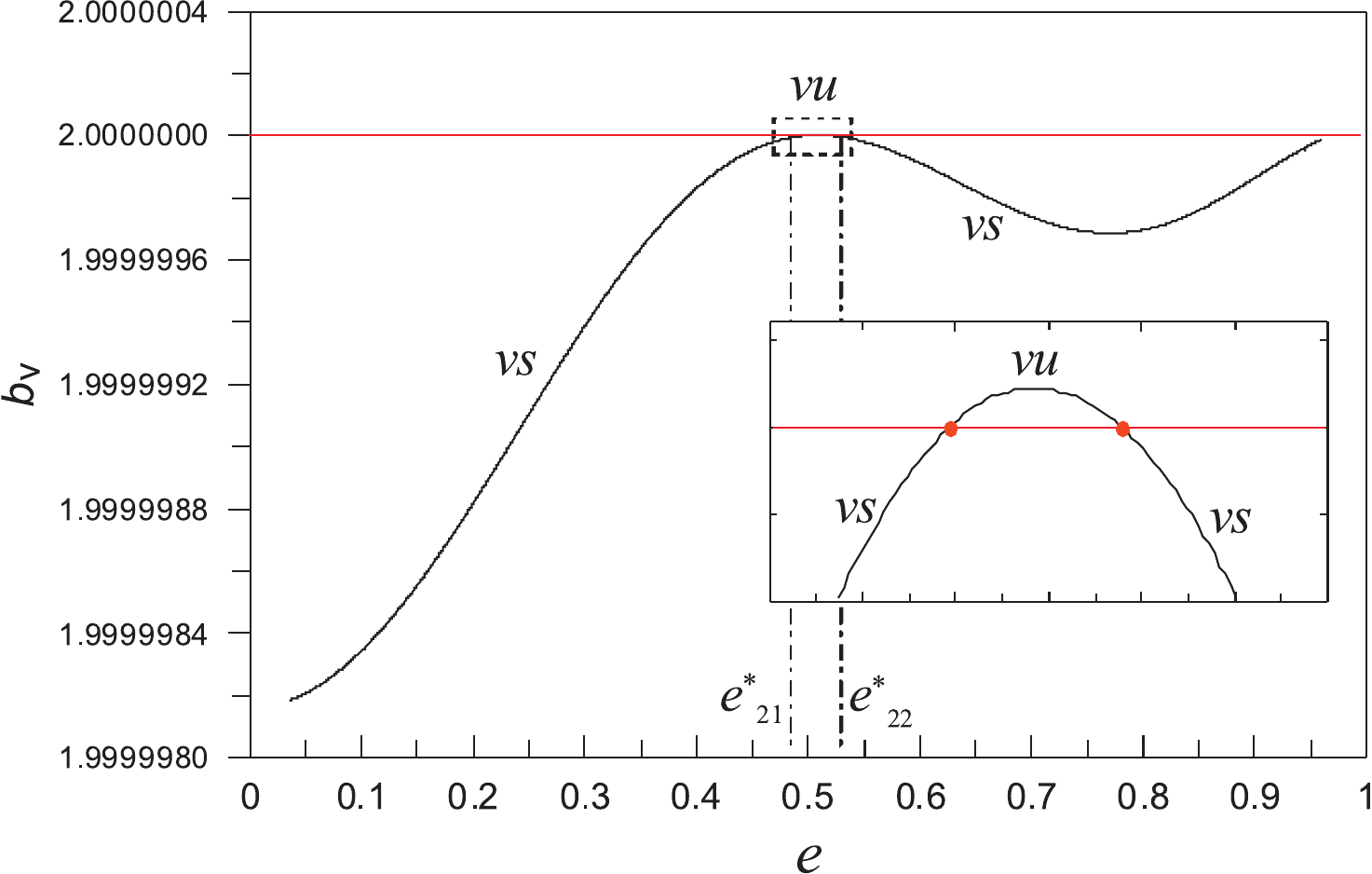} & \quad & \includegraphics[width=0.45\textwidth]{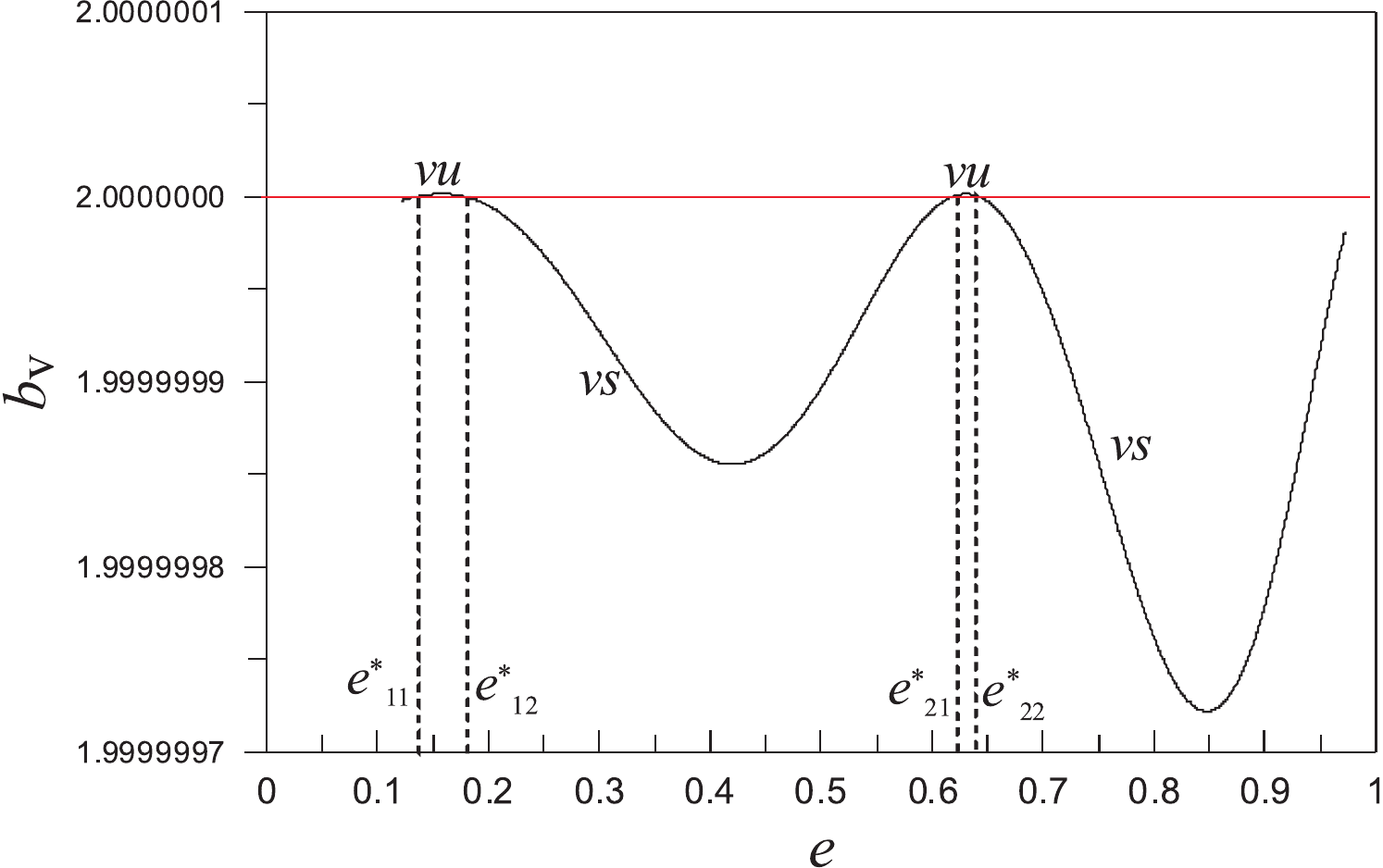} \\
\textnormal{(a)} & \quad & \textnormal{(b)}  
\end{array}$
\caption{The index $b_v$ of vertical stability along family $A_{II}$ for $\mu=5.15\cdot10^{-5}$  in {\bf a} $1:2$  and  {\bf b} $1:3$ exterior MMR. Vertical instability (vu) exists when $b_v>2$, i.e. in the eccentricity intervals ($e^*_{11},e^*_{12}$) and ($e^*_{21},e^*_{22}$)}
\label{FIGVSTAB}
\end{figure} 

We computed the index $b_v$ for all a.p.o. of family $A_{II}$. Figure \ref{FIGVSTAB}a presents the index $b_v$ along the $1:2$ resonant family, which is parametrized with respect to the eccentricity, $e$, of the orbits. We obtain that $b_v$ becomes larger than 2 between the eccentricity values $e^*_{21}$ and $e^*_{22}$. Between these neighbouring points, the orbits are vertically unstable. So we conclude that the majority of $1:2$ resonant orbits of $A_{II}$ are vertically stable and a pair of v.c.o. exists at eccentricities $e^*_{21}$ and $e^*_{22}\sim 0.5$.  
This result was also found by \cite{Markellos77}. However, for the resonances $1:3$, $1:4$ and $1:5$ we obtain two pairs of v.c.o., $(e^*_{11}, e^*_{12})$ and $(e^*_{21}, e^*_{22})$, as shown in Fig. \ref{FIGVSTAB}b. Table \ref{TABVCO} presents the eccentricity values of v.c.o. for the studied resonances. Now two small family segments of vertically unstable orbits exist but, similarly to the $1:2$ resonance, we conclude that the majority of orbits of $A_{II}$, which are horizontally stable, are also vertically stable. We should remark that as $e\rightarrow 1$ the precision of our computations degrades and the possibility that additional v.c.o. exist for very high eccentricities cannot be excluded. However, this goes beyond the scope of this study.        

\begin{table}
\caption{Eccentricity values for asymmetric v.c.o. ($\mu=0.000515$)} 
{
\begin{tabular}{cccccc}
\hline
MMR & $e^*_{11}$ & $e^*_{12}$ &  & $e^*_{21}$ & $e^*_{22}$ \\ 
\hline
$1:2$  &   -  &  -     & $\quad$ & 0.500 & 0.517 \\ 
$1:3$  & 0.137 & 0.179 & $\quad$ & 0.619 & 0.642 \\
$1:4$  & 0.273 & 0.318 & $\quad$ & 0.685 & 0.709 \\
$1:5$  & 0.366 & 0.409 & $\quad$ & 0.728 & 0.752 \\
\hline
\end{tabular}
}
\label{TABVCO}
\end{table}

\section{Spatial asymmetric periodic orbits}

\subsection{Computational approach} \label{CompuProc}
Let us denote an orbit by $\mathbf{X}=\mathbf{X}(t, \mathbf{X}_0)$, where $\mathbf{X}=(x,y,z,\dot x, \dot y, \dot z)$ is the position vector in phase space and $\mathbf{X}_0$ the initial condition vector. We can always consider initial conditions on the Poincar\'e surface of section $y=0$ and describe the orbit by the corresponding Poincar\'e map. A periodic orbit of initial conditions $\mathbf{X}_{00}$ and period $T$,
\begin{equation} \label{DEFAPO} 
\mathbf{X}(T; \mathbf{X}_{00})=\mathbf{X}_{00},\quad\quad \mathbf{X}_{00}=(x_{00},0,z_{00},\dot x_{00},\dot y_{00},\dot z_{00}),
\end{equation}
is represented by $k$ points of the Poincar\'e map, where $k$ is the multiplicity of the orbit. If $\dot x_{00}=\dot x(T/2)=0$ and $\dot z_{00}=\dot z(T/2)=0$, then the orbit is symmetric with respect to the $xz$-plane. If $\dot x_{00}=\dot x(T/2)=0$ and $z_{00}=z(T/2)=0$, the orbit is symmetric with respect to the $x$-axis \citep{henon73, kv05}. The above relations constitute also the periodic conditions which should be satisfied, in order for the orbit to be an s.p.o. For an a.p.o. all components of the initial state vector $\mathbf{X}_{00}$ (except $y_{00}$) are generally non-zero and the four periodic conditions
\begin{equation} \label{APOCON}
x(T;\mathbf{X}_{00})=x_{00},\quad  z(T;\mathbf{X}_{00})=z_{00},\quad \dot x(T;\mathbf{X}_{00})=\dot x_{00},\quad \dot y(T;\mathbf{X}_{00})=\dot y_{00},
\end{equation} 
should be satisfied. The condition $y(T;\mathbf{X}_{00})=0$ is always satisfied, due to the Poincar\'e map and $\dot z(T;\mathbf{X}_{00})=\dot z_{00}$, because of the Jacobi integral.      

Considering an a.p.o. (\ref{DEFAPO}) which satisfies the conditions  (\ref{APOCON}), we are looking for a periodic solution with $z_0=z_{00}+\delta z^*$, where $\delta z^*$ is fixed and sufficiently small.  We assume a solution with initial conditions $\mathbf{X}_{01}=\mathbf{X}_{00} + \delta \mathbf{X}$, where $\delta \mathbf{X} = (\delta x, 0, \delta z^*, \delta \dot x, \delta \dot y, \delta \dot z)$, that satisfies the periodicity conditions (\ref{APOCON_}), namely
\begin{equation} \label{APOCON_}
\begin{array}{lll}
x(T_k;\mathbf{X}_{00}+\delta \mathbf{X})=x_{00}+\delta x,& \quad & z(T_k;\mathbf{X}_{00}+\delta \mathbf{X})=z_{00}+\delta z^*, \\
\dot x(T_k;\mathbf{X}_{00}+\delta \mathbf{X})=\dot x_{00}+\delta \dot x, & \quad & \dot y(T_k;\mathbf{X}_{00}+\delta \mathbf{X})=\dot y_{00}+\delta \dot y. 
\end{array}
\end{equation}
In a first order approximation with respect to $\delta \mathbf{X}$, the differential corrections $\delta x$, $\delta \dot x$, $\delta \dot y$, and  $\delta \dot z$ can be computed by the equation
\begin{equation} \label{DIFFCORR}
\left ( \begin{array}{c} \delta x \\ \delta \dot x \\ \delta \dot y \\ \delta \dot z \end{array} \right ) = 
\left ( \begin{array}{cccc} 
\frac{\partial x}{\partial x_0}-1 & \frac{\partial x}{\partial \dot x_0} & \frac{\partial x}{\partial \dot y_0} & \frac{\partial x}{\partial \dot z_0} \\
\frac{\partial z}{\partial x_0} & \frac{\partial z}{\partial \dot x_0} & \frac{\partial z}{\partial \dot y_0} & \frac{\partial z}{\partial \dot z_0} \\
\frac{\partial \dot x}{\partial x_0} & \frac{\partial \dot x}{\partial \dot x_0}-1 & \frac{\partial \dot x}{\partial \dot y_0} & \frac{\partial \dot x}{\partial \dot z_0} \\
\frac{\partial \dot y}{\partial x_0} & \frac{\partial \dot y}{\partial \dot x_0} & \frac{\partial \dot y}{\partial \dot y_0} -1 & \frac{\partial \dot y}{\partial \dot z_0} 
\end{array} \right ) ^{-1}_0  
\left ( \begin{array}{c}  x(T_k;\mathbf{X}_0) - x_{00}  \\ z(T_k;\mathbf{X}_0) - z_{00} \\
\dot x(T_k;\mathbf{X}_0) - \dot x_{00}  \\ \dot y(T_k;\mathbf{X}_0) - \dot y_{00}
\end{array} \right ).
\end{equation}
where $\mathbf{X}_{0}=(x_{00},0,z_{00}+\delta z^*,\dot x_{00},\dot y_{00},\dot z_{00})$ and $T_k$ is the time of the predefined $k^{th}$ intersection with the surface of section. We repeat the procedure by replacing $\mathbf X_{00}$ with $\mathbf X_{01}$. If the procedure converges then $|\delta \mathbf{X}| \rightarrow 0$ and $T_k$ tends to the period of the new periodic orbit  $\mathbf X_{01}$. We stop the computations, when the orbit returns after $T_k$ to its origin with deviation $|\mathbf X_{01}-\mathbf X_{00}|\leq acc$, where we set $acc=10^{-11}$ or, if it is feasible, $acc=10^{-12}$.	Having computed the last orbit, the continuation process seeks for a new orbit with initial condition $z_{02}=z_{01}+\delta z^*$ e.t.c. Our computations showed that the above procedure is possible if we start from a v.c.o. of an asymmetric planar family, and, thus, the conjecture of Markellos is verified.  
   	
\begin{figure}
$
\begin{array}{ccc}
\includegraphics[width=0.45\textwidth]{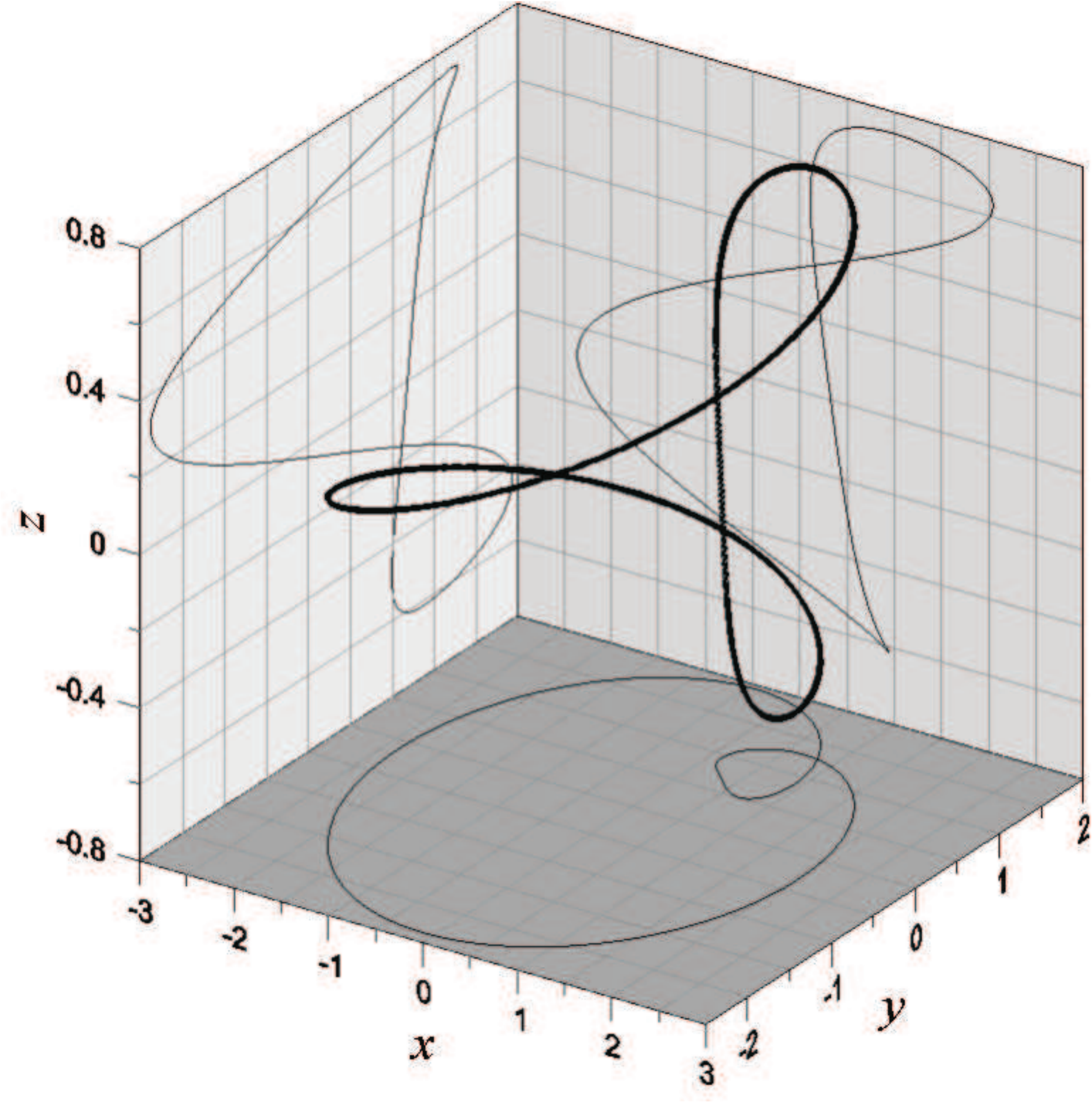} & \quad & \includegraphics[width=0.45\textwidth]{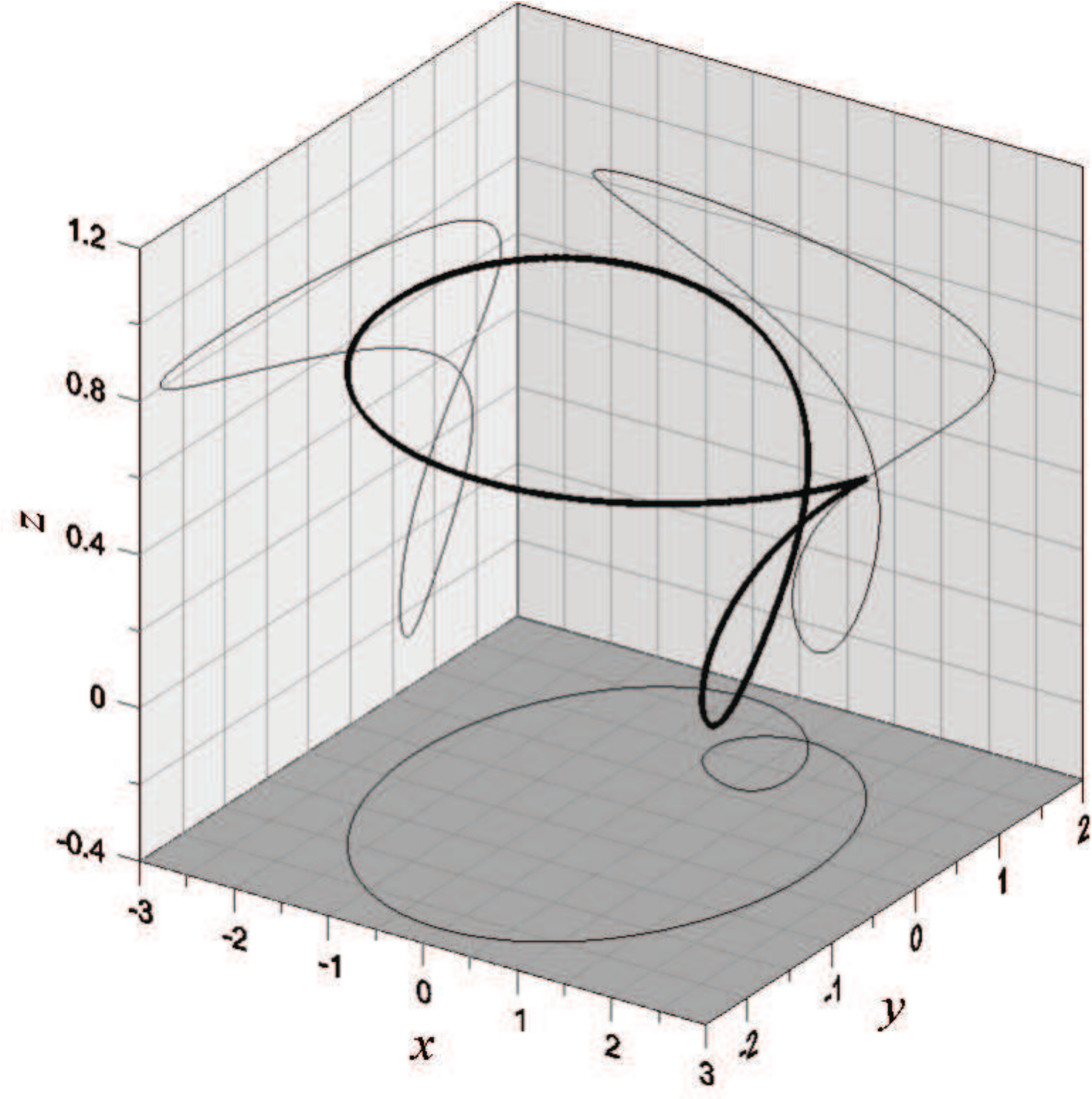} \\
\textnormal{(a)} & \quad & \textnormal{(b)}  
\end{array}$
\caption{Spatial a.p.o. (thick curve) in $1:2$ MMR with Neptune  {\bf a } {\em orbit 1} with initial conditions $x_0=2.021351199750$, $y_0=0$, $z_0=0.65848014126$, $\dot x_0=0.29098336455$, $\dot y_0=-1.56229011545$, $\dot z_0=-0.12410410620$  {\bf b } {\em orbit 2} with $x_0=2.05314832761$, $y_0=0$, $z_0=0.57686260649$, $\dot x_0=0.19391519007$, $\dot y_0=-1.56371158918$, $\dot z_0=0.17516988888$. The projections of the orbits on the planes $xy$, $xz$ and $yz$ are also shown with thinner curves} 
\label{FigORBITS3D}
\end{figure} 

In Fig. \ref{FigORBITS3D}, we present two samples of $1:2$ a.p.o. in three dimensions which belong to different families as we will show in the following sections. The initial conditions in the rotating frame are given in the caption. The orbits are almost Keplerian ellipses in the inertial frame with the orbit in panel (a) (called {\em orbit 1}) corresponding to orbital elements $a=1.58727$, $e=0.526$, $i=30^\circ$, $\omega=351^\circ$, $\Omega=214^\circ$ and $M=216^\circ$, while the orbit in panel (b) (called {\em orbit 2}) corresponds to $a=1.58725$, $e=0.524$, $i=20.5^\circ$, $\omega=259^\circ$, $\Omega=311^\circ$ and $M=221^\circ$. Here, $\omega$ is the argument of pericentre and $\Omega$ is the longitude of the ascending node.     				
				
\subsection{Stability of a.p.o.}
\begin{figure}
$
\begin{array}{ccc}
\includegraphics[width=0.45\textwidth]{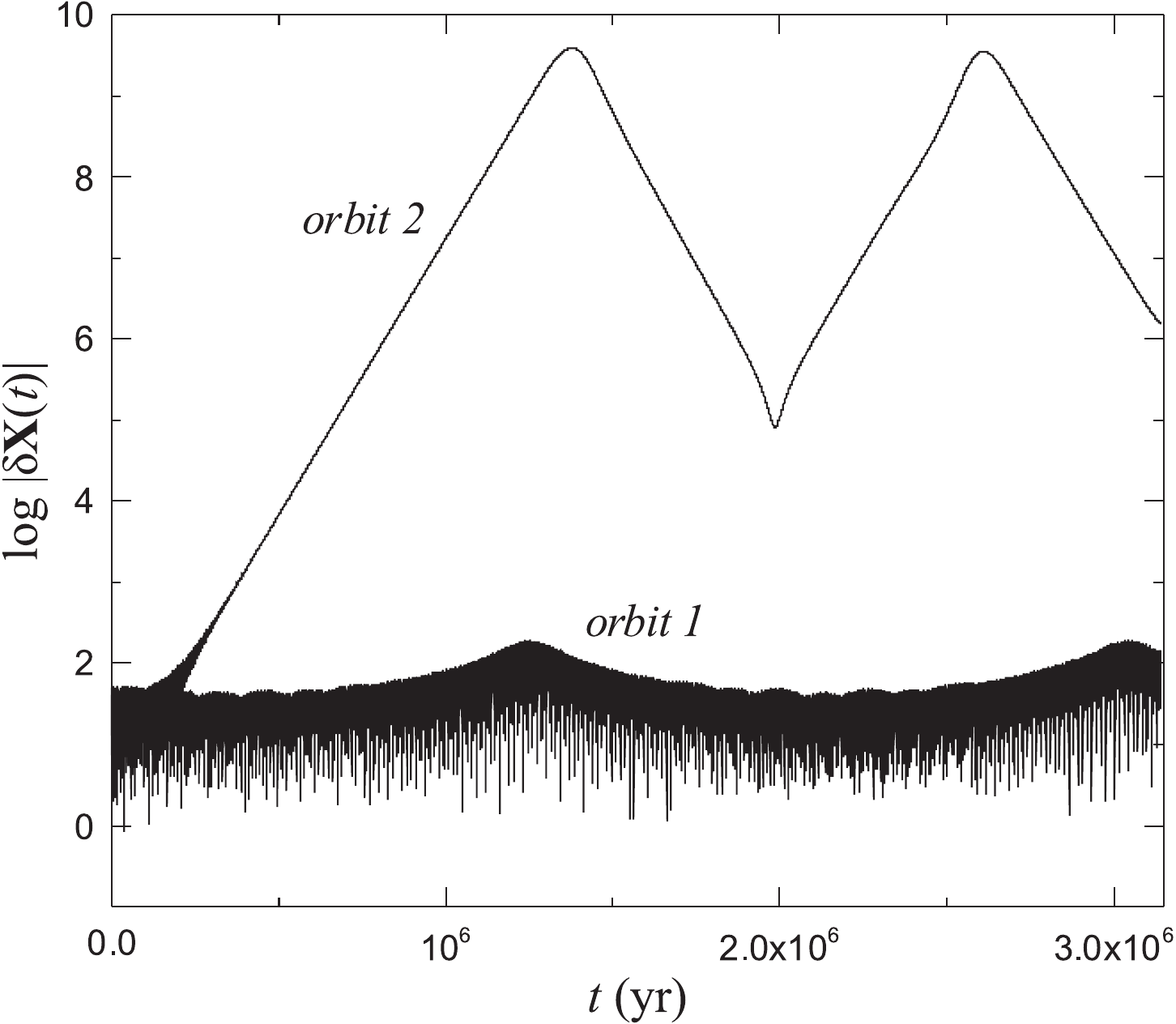} & \quad & \includegraphics[width=0.45\textwidth]{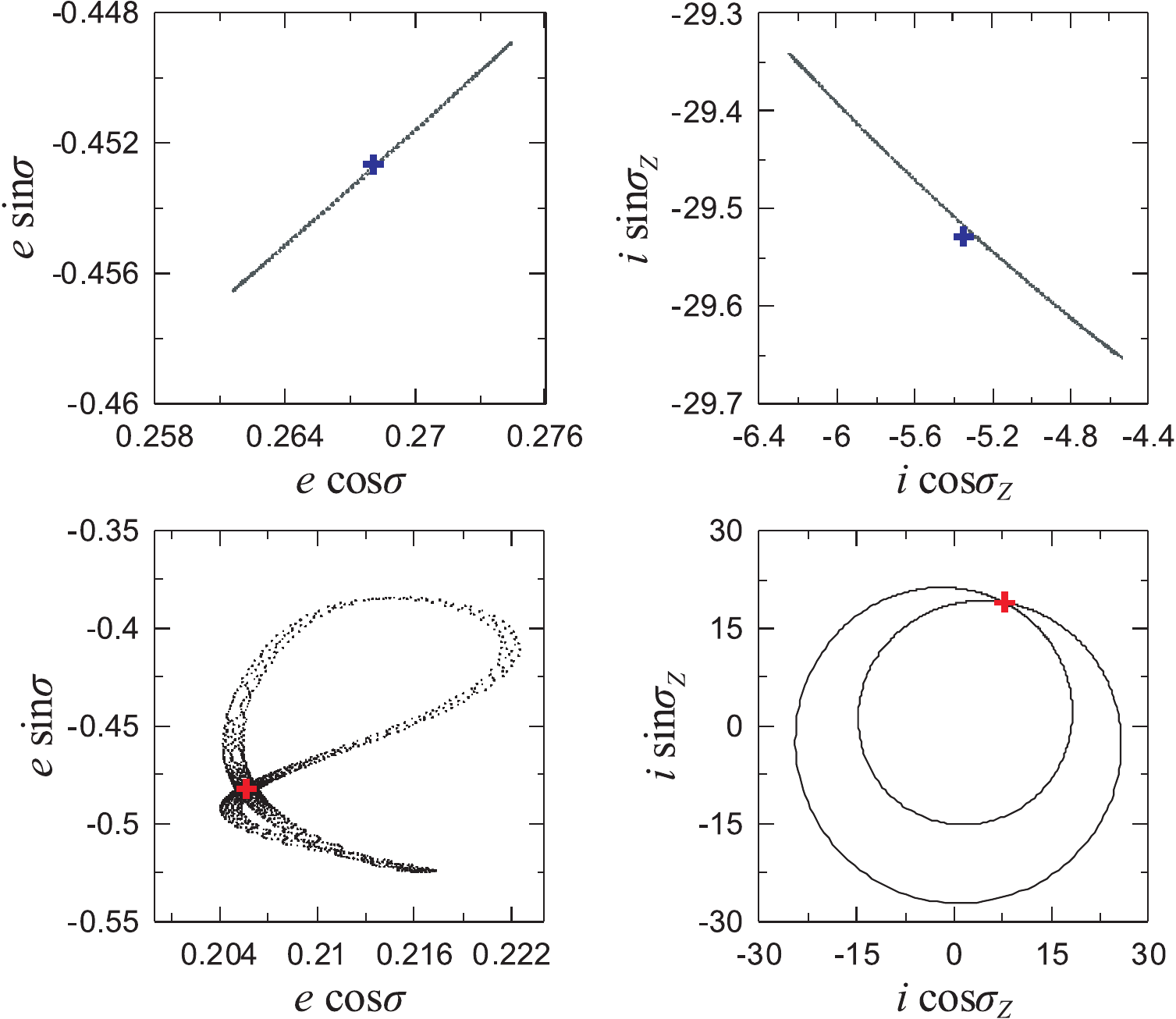} \\
\textnormal{(a)} & \quad & \textnormal{(b)}  
\end{array}
$
\caption{{\bf a} The evolution of the norm of the deviation $\delta \mathbf{X}(t)$ along {\em orbit 1} and {\em orbit 2} of Fig. \ref{FigORBITS3D}. {\bf b} Projections of Poincar\'e surfaces of section of orbits starting close to the orbits {\em 1} (upper panels) and {\em 2} (bottom panels) indicated by the cross}   
\label{FIGSTABSAMPLE}
\end{figure}

We can study the linear stability of a.p.o. through the variational equations of the system and the monodromy matrix, which is computed for a full period of the particular periodic solution. We follow the procedure given by \cite{Broucke69} and compute the stability indices. Since the mass of Neptune, which is used in our computations, is relatively very small ($\mu<<1$), the stability indices are very close to their critical values which correspond to the unperturbed case ($\mu=0$).
So due to limited computational accuracy, in many cases the results for linear stability are ambiguous. In order to clarify the situation, we compute the deviation vector $\delta \mathbf{X}(t)$ by integrating the variational equations for a long time interval, $0\leq t \leq t_{max}$, and by assuming as stability index the quantity
\begin{equation}
d=\textnormal{max} \left \{ ||\delta \mathbf{X}(t)||; \;0\leq t\leq t_{max}; \;||\delta \mathbf{X}(0)||=1 \right \} .
\end{equation}
In case of a stable periodic orbit, $d$ should take small values relatively to the values obtained in case of an unstable periodic orbit. An example is presented in Fig. \ref{FIGSTABSAMPLE}a, where we present the evolution of $||\delta \mathbf{X}(t)||$ along the {\em orbit 1} and {\em orbit 2} of Fig. \ref{FigORBITS3D} for $t_{max}\approx 3My$. For the index $d$ we obtain approximately the values $10^{2.1}$ and $10^{9.7}$, respectively, and the {\em orbit 1} is classified as stable, while {\em orbit 2} as unstable. 

The stability character may be also revealed by presenting the orbits on projections of the 4-dimensional surfaces of section $y=0$. In Fig. \ref{FIGSTABSAMPLE}b, we present projections on the planes $(e \cos \sigma,\, e \sin \sigma)$ and $(i \cos \sigma_z,\, i \sin \sigma_z)$, where $\sigma_z=2\lambda'-4\,\lambda+2\Omega$ is the resonant argument for the inclined $1:2$ exterior resonance. The upper panels correspond to an orbit with initial conditions as those of the stable a.p.o. {\em orbit 1} when a deviation to $z_0$ of order $10^{-3}$ is added. We see that this orbit remains close to the periodic orbit for very long times. The bottom panels correspond to an orbit close to the unstable periodic orbit ({\em orbit 2}). Now we used as initial conditions the ones of {\em orbit 2} but we added a very small deviation of order $10^{-10}$. In a long time integration  interval ($t_{max}\approx 100My$) the orbit diverges significantly from its original position and we find a ``saddle'' configuration. Chaos certainly exists but it should be confined in a very narrow strip around the saddle configuration. Hence, the original periodic orbit is weakly unstable and it is not associated with strongly irregular behaviour. 

\subsection{Orbital evolution near a.p.o.}
\begin{figure}
$
\begin{array}{ccccc}
\includegraphics[height=8.5cm]{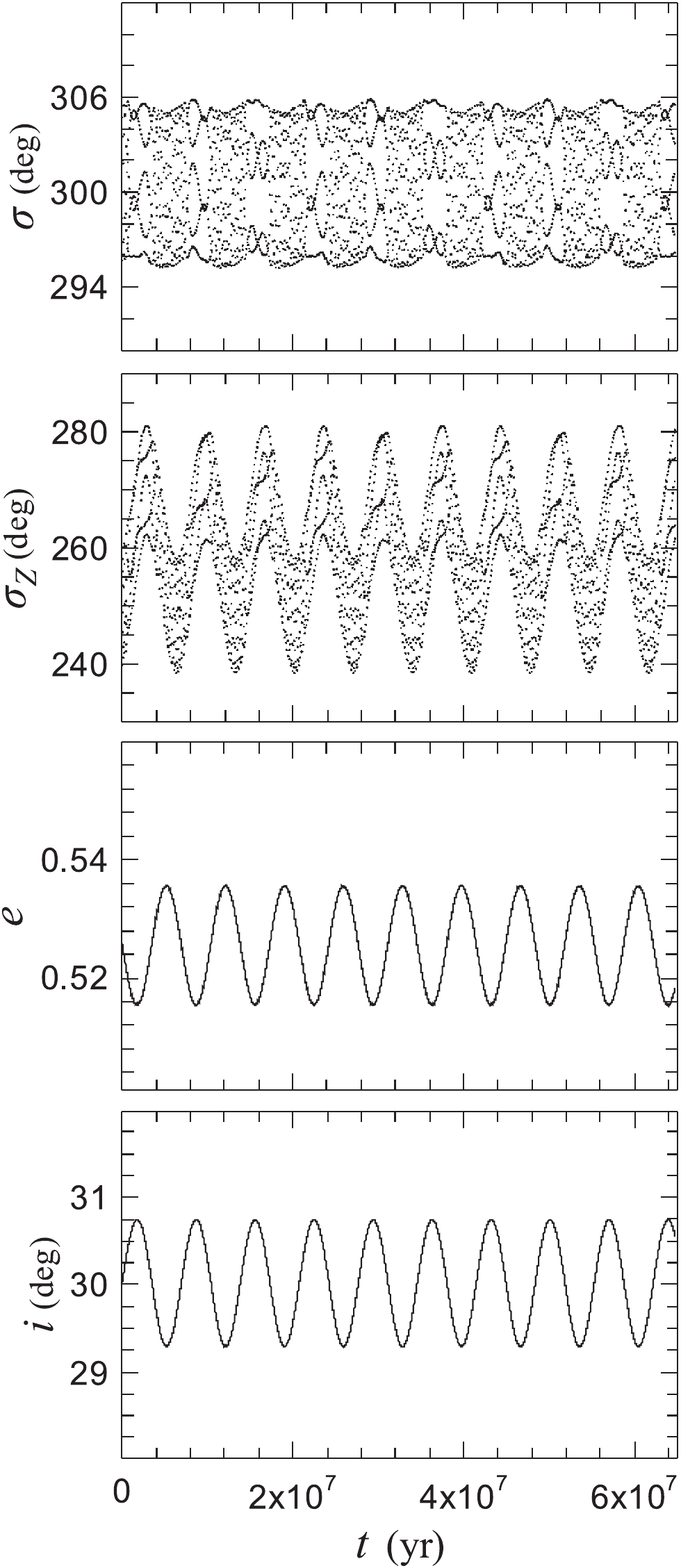} & \quad & \includegraphics[height=8.5cm]{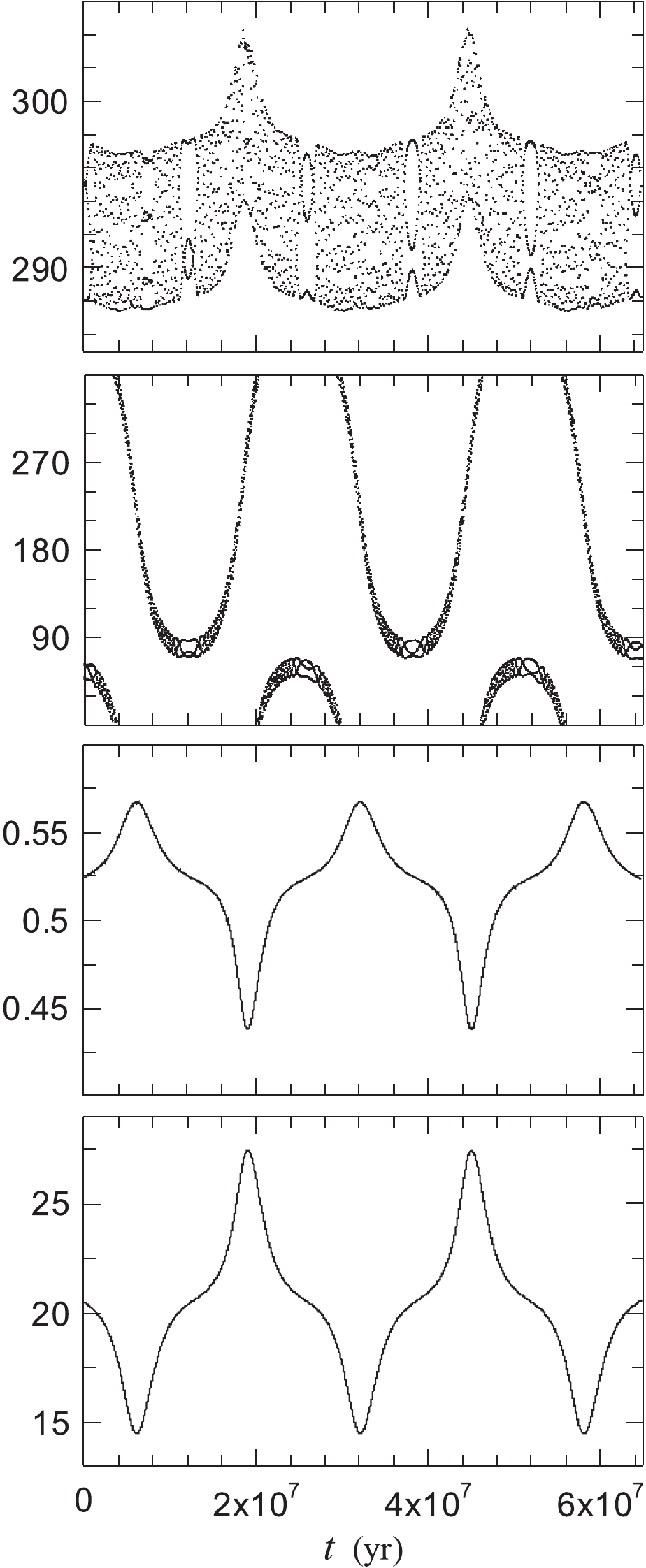} & \quad & \includegraphics[height=8.5cm]{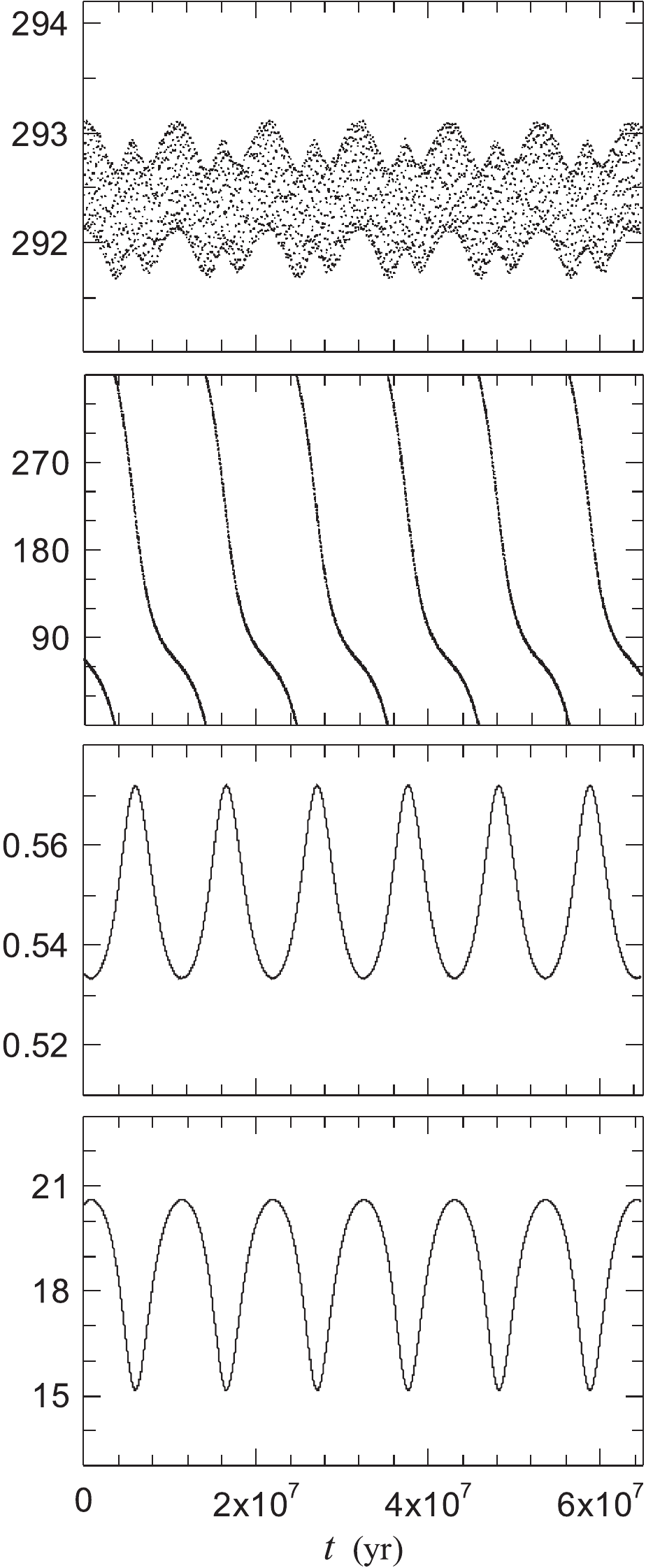}\\
\textnormal{(a)} & \quad & \textnormal{(b)} & \quad & \textnormal{(c)}  
\end{array}
$ 
\caption{ Evolution of $\sigma$, $\sigma_Z$, $e$ and $i$ along an orbit {\bf a} initially deviated from {\em orbit 1} by $\Delta \omega = 5^\circ$ {\bf b} initially deviated from {\em orbit 2} by $\Delta \omega = 5^\circ$ {\bf c} initially deviated from {\em orbit 2} by $\Delta e = 5^\circ$}   
\label{FIGLIBROT}
\end{figure} 
According to the conditions of KAM theory, stable periodic orbits should be foliated by invariant tori associated with quasiperiodic motion \citep{Berry1978}. Small deviation of initial conditions from those of the stable periodic orbit, generally, leads to orbits where the almost constant quantities, such as the semimajor axis, the eccentricity and the inclination, oscillate with small amplitudes. Instead, in the neighbourhood of an unstable periodic orbit, the existence of unstable manifolds may cause large variations, mainly to the eccentricity and inclination. Considering an $1:p$ MMR in the spatial CRTBP, we can define a set of resonant angles of the form 
$$
\bar \sigma = k_1(\lambda'-p \lambda)+k_2 \varpi + k_3 \Omega,
$$      
where the integers $k_i$ and $p$ must satisfy the D'Alembert rules \citep{murraydermott}. The presence of an MMR is indicated by the libration of at least one resonant angle. However, in the neigbourhood of the stable exact resonance, which is given by a stable periodic orbit, all resonant angles should librate. We remind that for asymmetric periodic orbits such librations take place around values other than $0$ or $\pi$. 

We support the above argument by considering orbits close to the a.p.o. presented in Fig. \ref{FigORBITS3D} and in Fig. \ref{FIGLIBROT} we present the time evolution along these orbits of the resonant angles $\sigma=\lambda'-2\,\lambda+\varpi$ and $\sigma_Z=2(\lambda'-2\,\lambda)+2\Omega$ and the eccentricity and the inclination. In panel a, we consider the orbit deviating initially from the stable {\em orbit 1} by $\Delta \omega=5^\circ$. Both $\sigma$ and $\sigma_Z$ librate while $e$ and $i$ show regular, anti-correlated slow oscillations due to the conservation of the angular momentum. The libration of both $\sigma$ and $\sigma_Z$ imply the libration of $\omega$ indicating the presence of Kozai resonance. In panel b, we consider an orbit with initial conditions close to the unstable {\em orbit 2} with deviation $\Delta \omega=5^\circ$. The angle $\sigma$ librates in this case too, but it shows an apparent slow oscillation of period $\sim 24$Myr. This periodic variation, which is the main period of the slow oscillation of $e$ and $i$, characterizes also the evolution of the angle $\sigma_Z$, which librates with an amplitude of $\sim 360^\circ$. Nevertheless, we can find orbits very close to the unstable {\em orbit 2} along which $\sigma_Z$ rotates. An example is shown in panel c, where we consider an initial deviation from {\em orbit 2} by $\Delta e=0.01$. The existence of both rotations and librations of $\sigma_Z$ close to the unstable orbit indicates the existence of a separatrix manifold associated with the unstable a.p.o.

\section{Families of a.p.o.}

\begin{figure}
$
\begin{array}{ccc}
\includegraphics[width=0.41\textwidth]{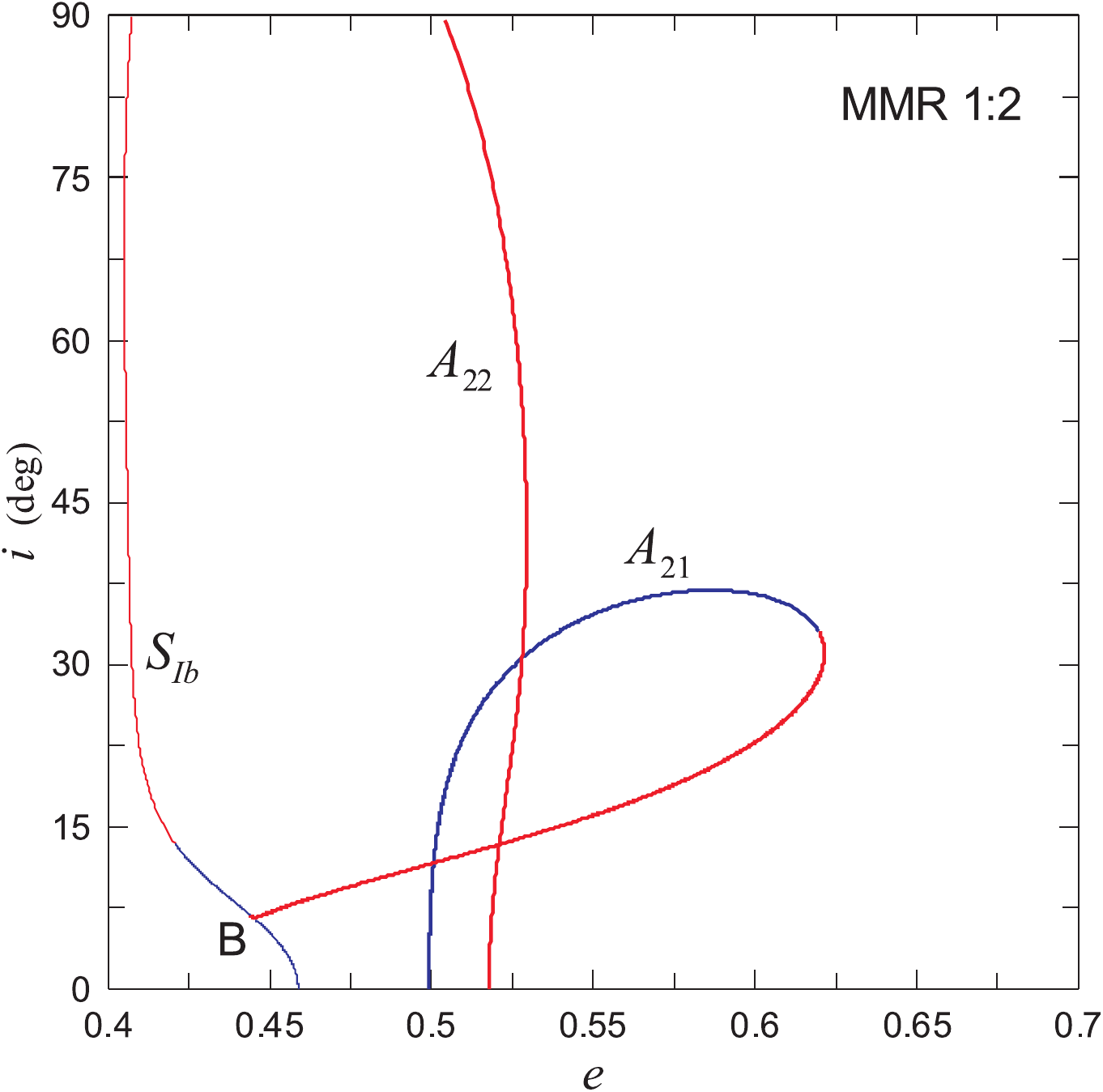} & \quad & \includegraphics[width=0.47\textwidth]{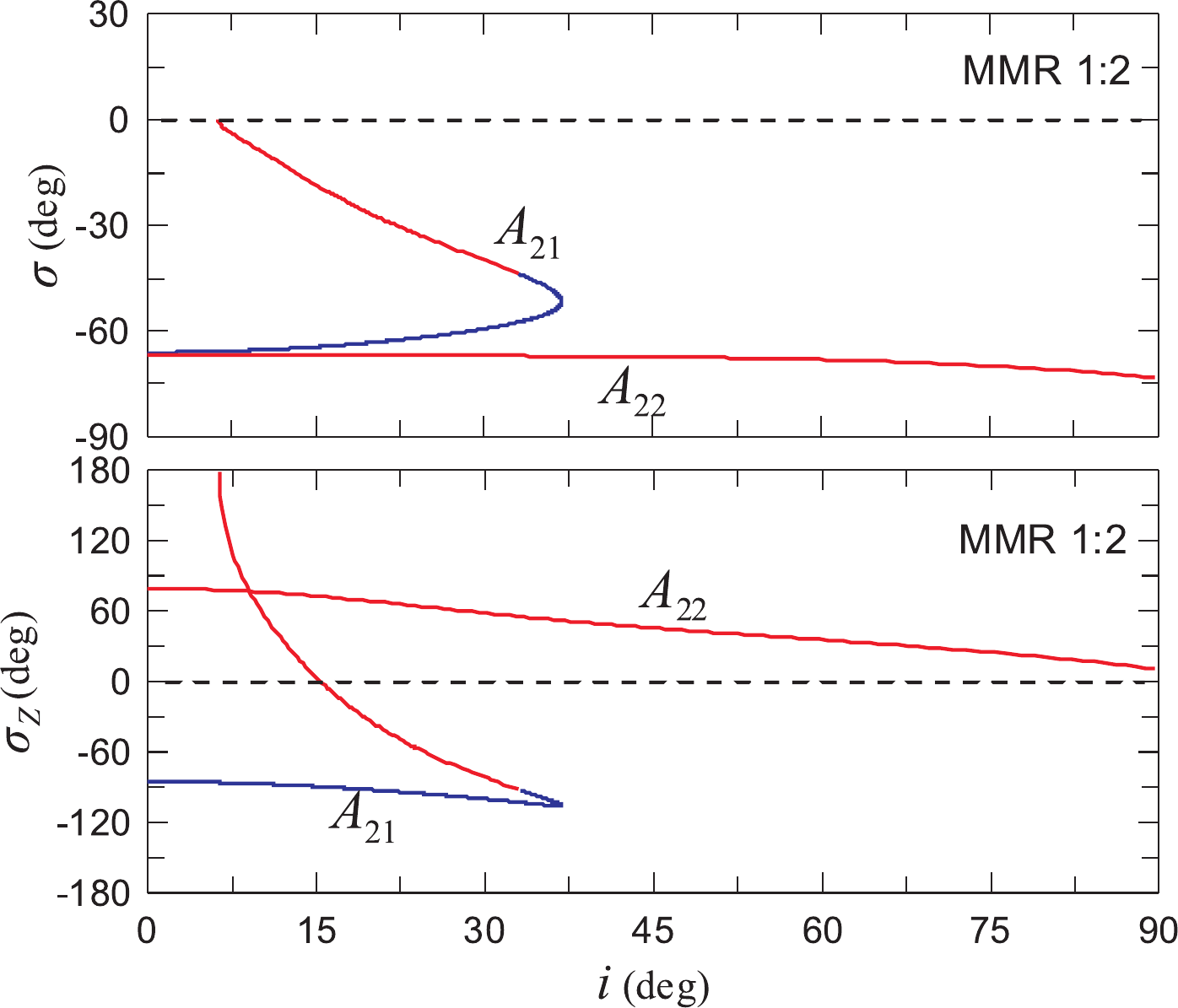} \\
\textnormal{(a)} & \quad & \textnormal{(b)}  
\end{array}
$
\caption{ Families of spatial asymmetric periodic orbits for the $1:2$ MMR {\bf a} Projection of the characteristic curves of periodic solutions on the $e-i$ plane. {\bf b} The values of the resonant angles $\sigma$ and $\sigma_z$ along the families (the inclination $i$ is used as the family parameter). Blue (red) line segments correspond to stable (unstable) solutions}   
\label{FIGAFAM12}
\end{figure}

Following the computational approach described in Sect. \ref{CompuProc} and starting from the v.c.o. of the planar asymmetric family $A_{II}$ we construct family segments of a.p.o. for the $1:p$, $p=2,...,5$ MMR. We denote these families by $A_{ij}$, where the indices refer to the v.c.o. of eccentricity $e_{ij}$ as given in Table \ref{TABVCO}. The families are projected on the plane $e-i$ and we compute also the variation of the resonant angle centres, along the families. 

Figure \ref{FIGAFAM12} presents families of the $1:2$ resonance. Family $A_{22}$ which starts from the plane with eccentricity $e=0.517$ continues up to very large inclinations, above $i=90^\circ$ where the orbits become retrograde. All orbits of $A_{22}$ are unstable. Family $A_{21}$ consists of stable orbits. It starts from the plane at $e=0.5$ and extends up to $i=37^\circ$, where $e=0.59$. The eccentricity reaches its maximum value at $e=0.62$ where the orbits turn to become unstable. Finally, $A_{21}$ terminates at a symmetric periodic orbit of the symmetric family $S_{Ib}$. This family bifurcates from the planar family $I_b$ at $e=0.456$. The family $S_{Ib}$ starts from the plane having stable orbits and at $B$, where $A_{21}$ terminates, it becomes unstable. However, it remains unstable only for a short inclination interval, about $6.5^\circ<i<6.8^\circ$, and then becomes stable again up to $i\approx 13^\circ$. From the variation of the resonant centres (panel b) we observe that at the bifurcation point $B$ we have $\sigma=0^\circ$ and $\sigma_Z=180^\circ$. These values of resonant angles characterize all symmetric orbits of the family $S_{Ib}$.   

\begin{figure}
$
\begin{array}{ccc}
\includegraphics[width=0.41\textwidth]{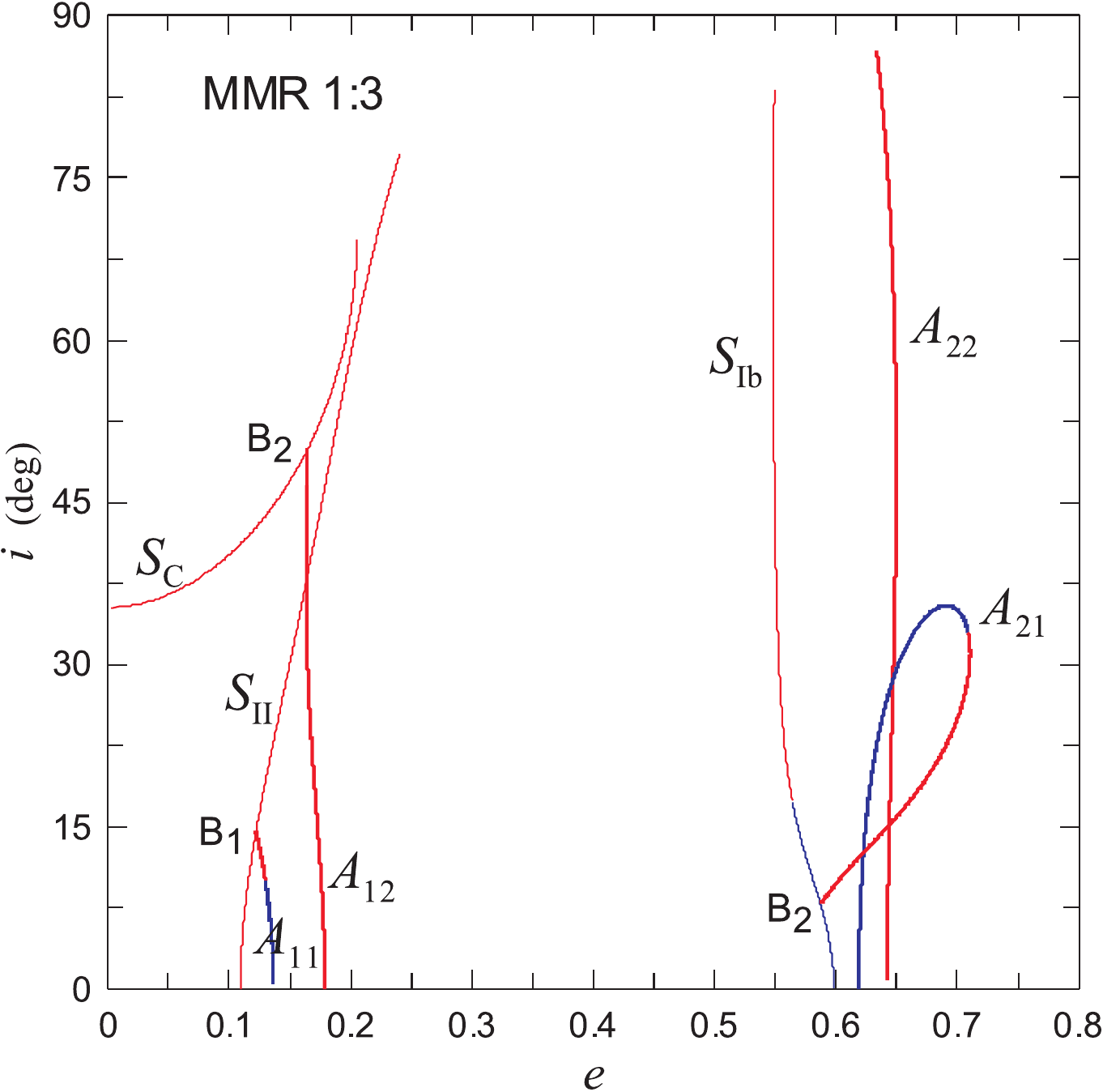} & \quad & \includegraphics[width=0.47\textwidth]{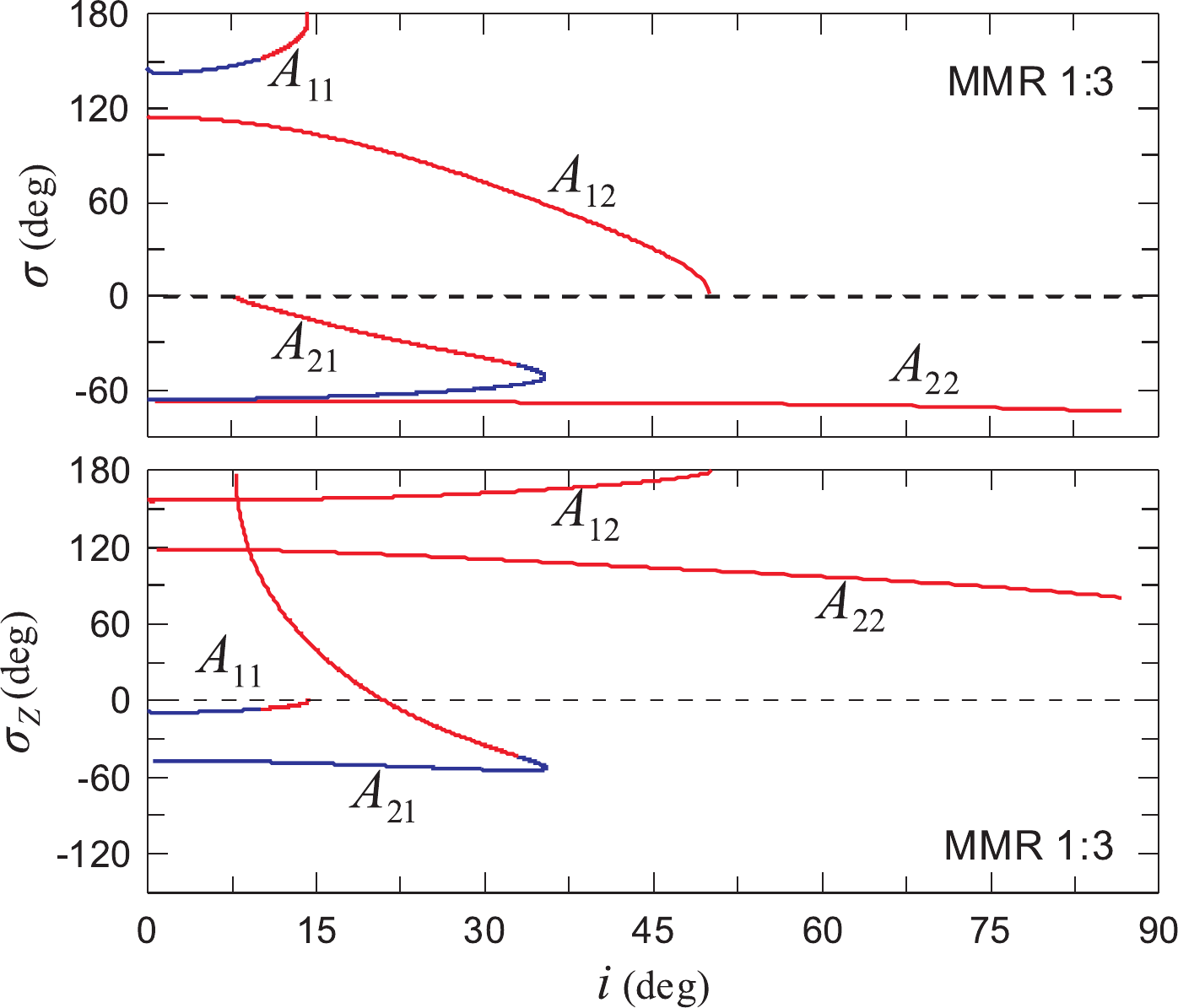} \\
\textnormal{(a)} & \quad & \textnormal{(b)}  
\end{array}
$
\caption{ Families of spatial asymmetric periodic orbits for the $1:3$ MMR (presentation as in Fig. \ref{FIGAFAM12})}   
\label{FIGAFAM13}
\end{figure}

The $1:3$ MMR of the planar case has four asymmetric v.c.o. and the generated asymmetric families are presented in Fig. \ref{FIGAFAM13}. Family $A_{11}$ starts with stable orbits, it becomes unstable at $i=10^\circ$ and at  $i=14.5^\circ$ terminates at the bifurcation point $B_1$, which belongs also to the symmetric family $S_{II}$ that is generated by a v.c.o. of the planar family $II$. This v.c.o. belongs to the unstable segment of the family and the generated symmetric family $S_{II}$ is unstable with $\sigma=180^\circ$ and $\sigma_Z=0^\circ$. Family $A_{12}$ consists entirely of unstable orbits and terminates at the symmetric family $S_c$, which bifurcates from an inclined circular orbit with $i\approx 40^\circ$. Families $A_{21}$ and $A_{22}$ show a similar structure as in $1:2$ MMR.

\begin{figure}
$
\begin{array}{ccc}
\includegraphics[width=0.41\textwidth]{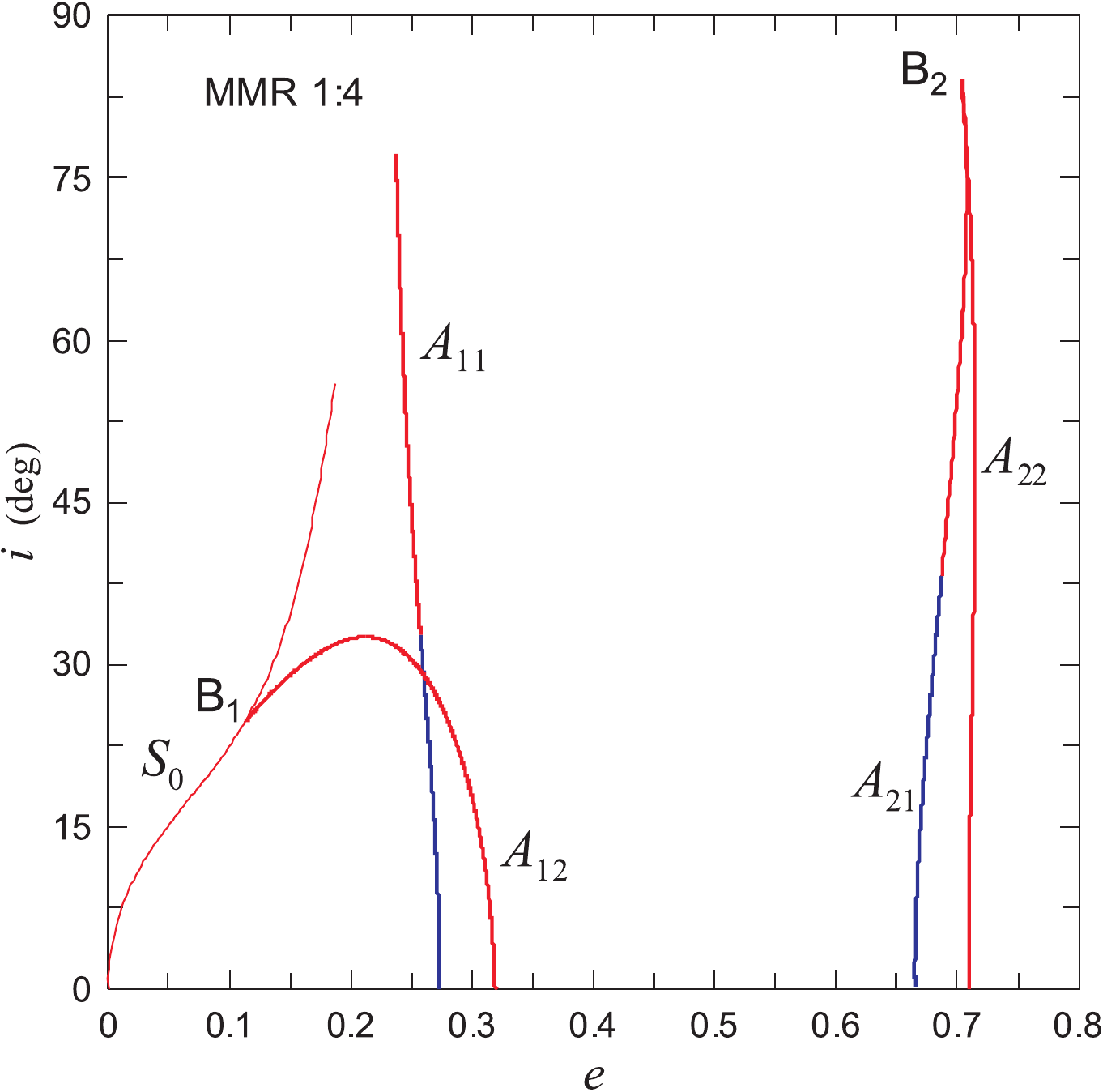} & \quad & \includegraphics[width=0.47\textwidth]{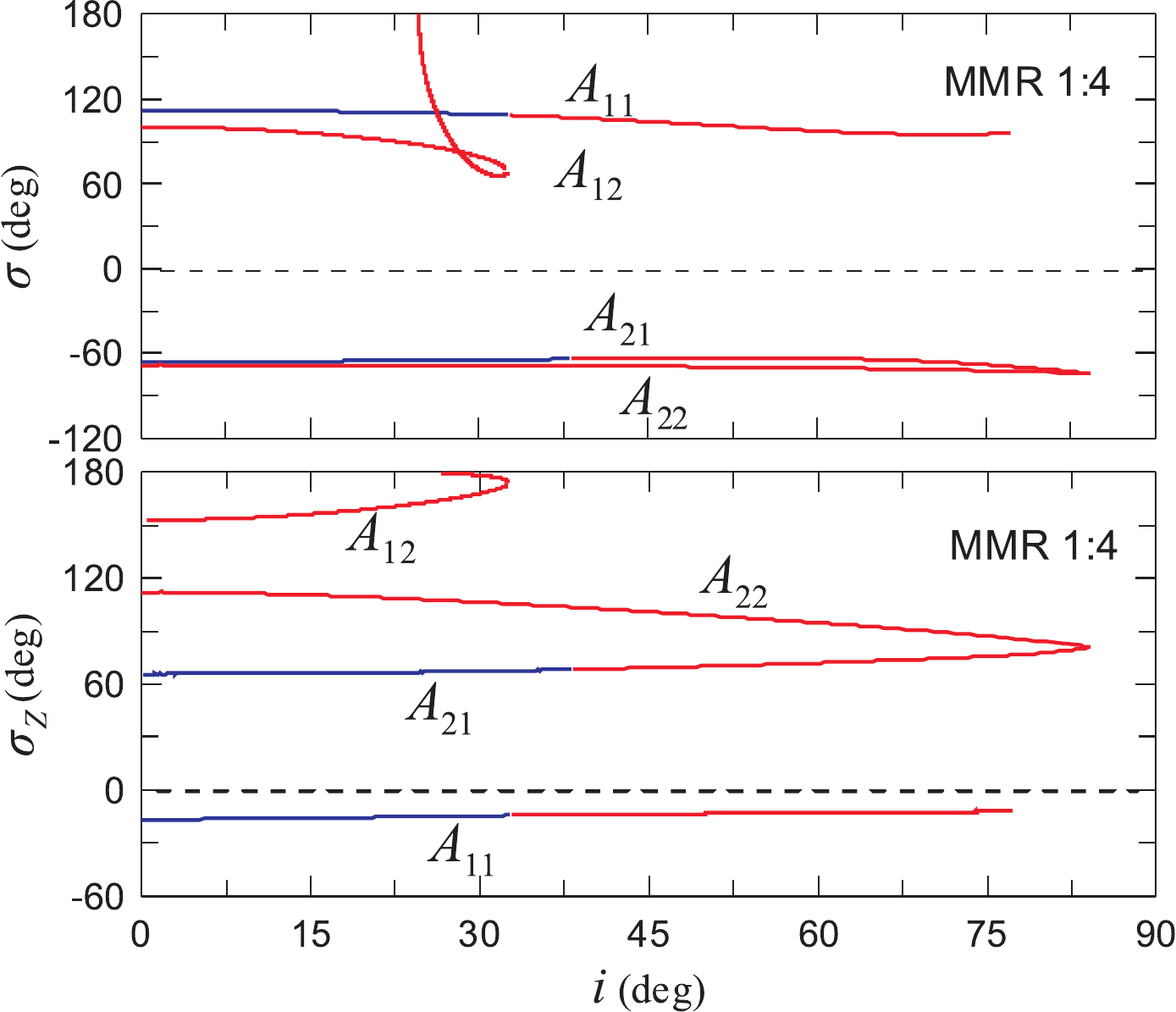} \\
\textnormal{(a)} & \quad & \textnormal{(b)}  
\end{array}
$
\caption{ Families of spatial asymmetric periodic orbits for the $1:4$ MMR (presentation as in Fig. \ref{FIGAFAM12})}   
\label{FIGAFAM14}
\end{figure} 

\begin{figure}
$
\begin{array}{ccc}
\includegraphics[width=0.41\textwidth]{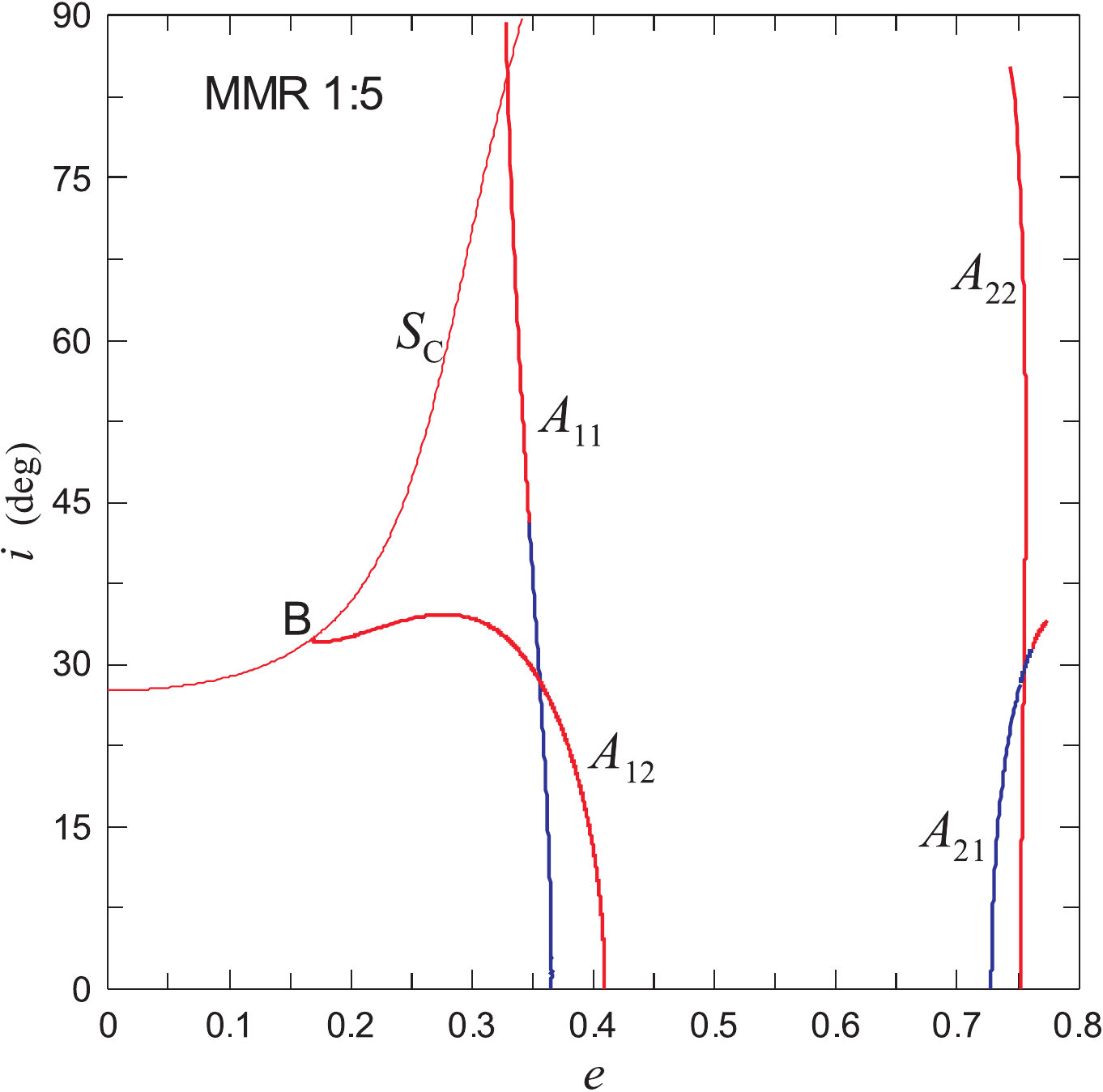} & \quad & \includegraphics[width=0.47\textwidth]{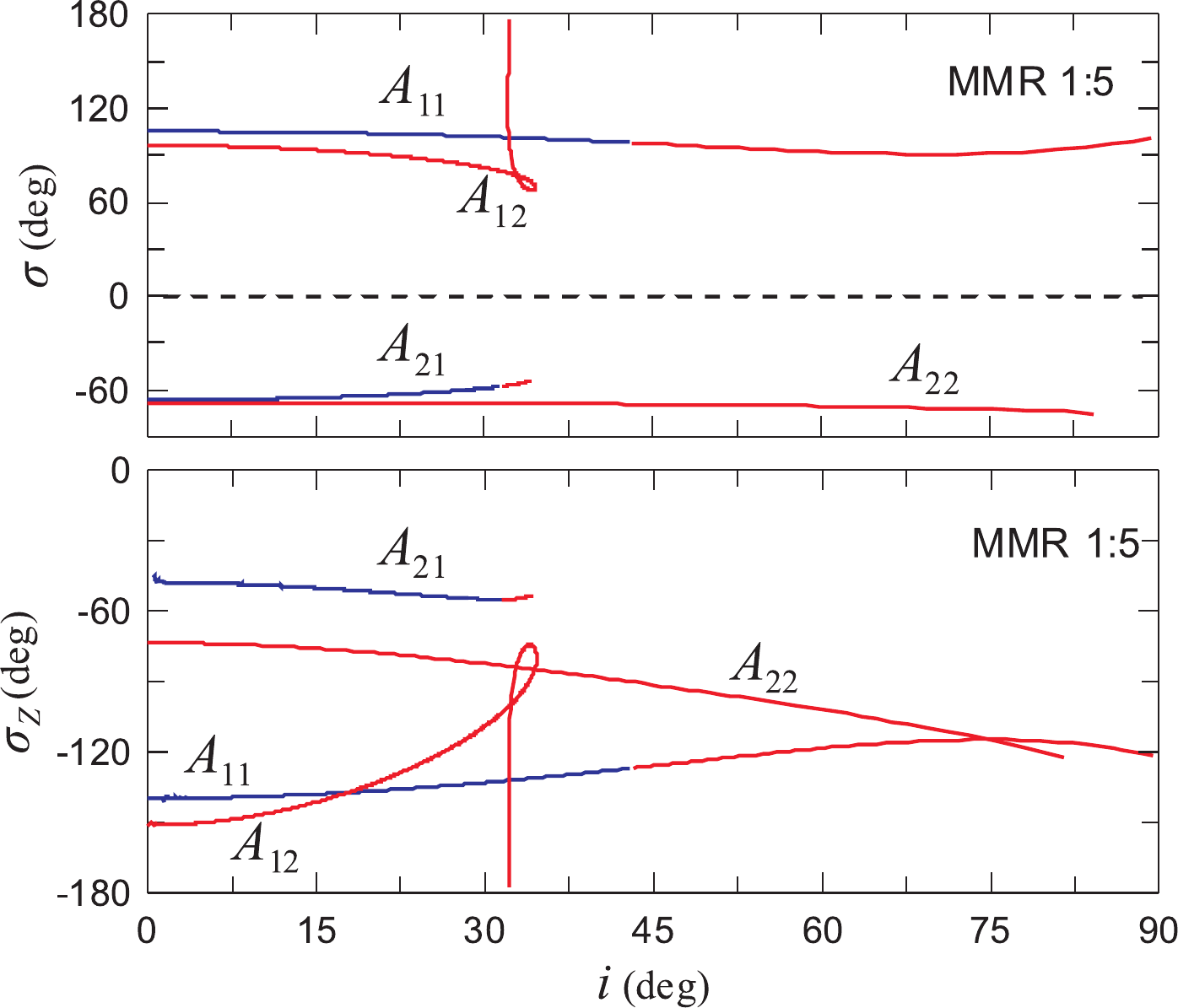} \\
\textnormal{(a)} & \quad & \textnormal{(b)}  
\end{array}
$
\caption{ Families of spatial asymmetric periodic orbits for the $1:5$ MMR (presentation as in Fig. \ref{FIGAFAM12})}   
\label{FIGAFAM15}
\end{figure} 

From the four asymmetric v.c.o. of the $1:4$ MMR we also obtained by continuation the four asymmetric families $A_{ij}$ presented in Fig. \ref{FIGAFAM14}. Family $A_{11}$ is stable up to $i=33^\circ$ and continues up to highly inclined orbits. Family $A_{12}$ is unstable and, as shown in the plot, it terminates to a symmetric orbit (point $B_1$) with $\sigma=\sigma_Z=180^\circ$. This orbit belongs to the symmetric family $S_0$, which is generated by a v.c.o. of the circular planar family $C$ close to the $3:1$ resonance. At this domain family $C$ is unstable and the generated family $S_C$ is also unstable. Family $A_{21}$ starts with stable orbits and the stability is preserved up to $i=38^\circ$. Then, the orbits become unstable and the family terminates by joining the family $A_{22}$ at $i=84^\circ$. Hence, a bridge is formed by the two families. 

The four $1:5$ resonant asymmetric families are presented in Fig. \ref{FIGAFAM15}.  Family $A_{11}$ is stable up to $i=43^\circ$ and continues up to large inclinations and unstable retrograde orbits. Family $A_{12}$ is unstable and terminates at a symmetric orbit that belongs to the unstable family $S_C$ which, similarly to the $1:3$ MMR, is generated by a circular inclined orbit with $i=28^\circ$. Family $A_{21}$ is stable up to $i=31^\circ$ where it becomes unstable at $e\approx 0.75$. For higher eccentricities, continuation becomes very slow and our algorithm has trouble determining the family's physical termination. Family $A_{22}$, as in previous MMRs, is unstable and extends up to high inclination values.

\section{Asymmetric librations of TNOs}

In this section, we extend our study by employing numerical integrations of the equations of motion of real solar system bodies as well as fictitious particles, in a model accounting for the gravitational perturbations exerted by the fully interacting four major planets (Jupiter, Saturn, Uranus and Neptune). All integrations were made using the {\em Mercury} software package \citep{Chambers99}, while initial conditions for the real bodies were taken from the AstDyS database\footnote{http://hamilton.dm.unipi.it}. 

We first tested the stability of orbits in the vicinity of the families of a.p.o. found in the previous section, for times equivalent to several My. The initial conditions were taken by properly rotating the orientation angles of the a.p.o. with respect to Neptune's orbit and position at the chosen epoch. The purpose of this experiment is to see if these 'stable orbits' would still remain stable under the additional perturbations. As expected, most orbits start off librating with very small amplitude about the central values found in the CRTBP but, after a few My, turn to circulation and chaos manifests itself. An exception to this rule is the $1:2$ MMR with Neptune, where a sizeable libration island can be found to host real TNOs. In this case, the a.p.o. (translated to the four-planets system) indeed remained stable for hundreds of My.

\begin{figure}
$
\begin{array}{ccc}
\includegraphics[width=0.45\textwidth]{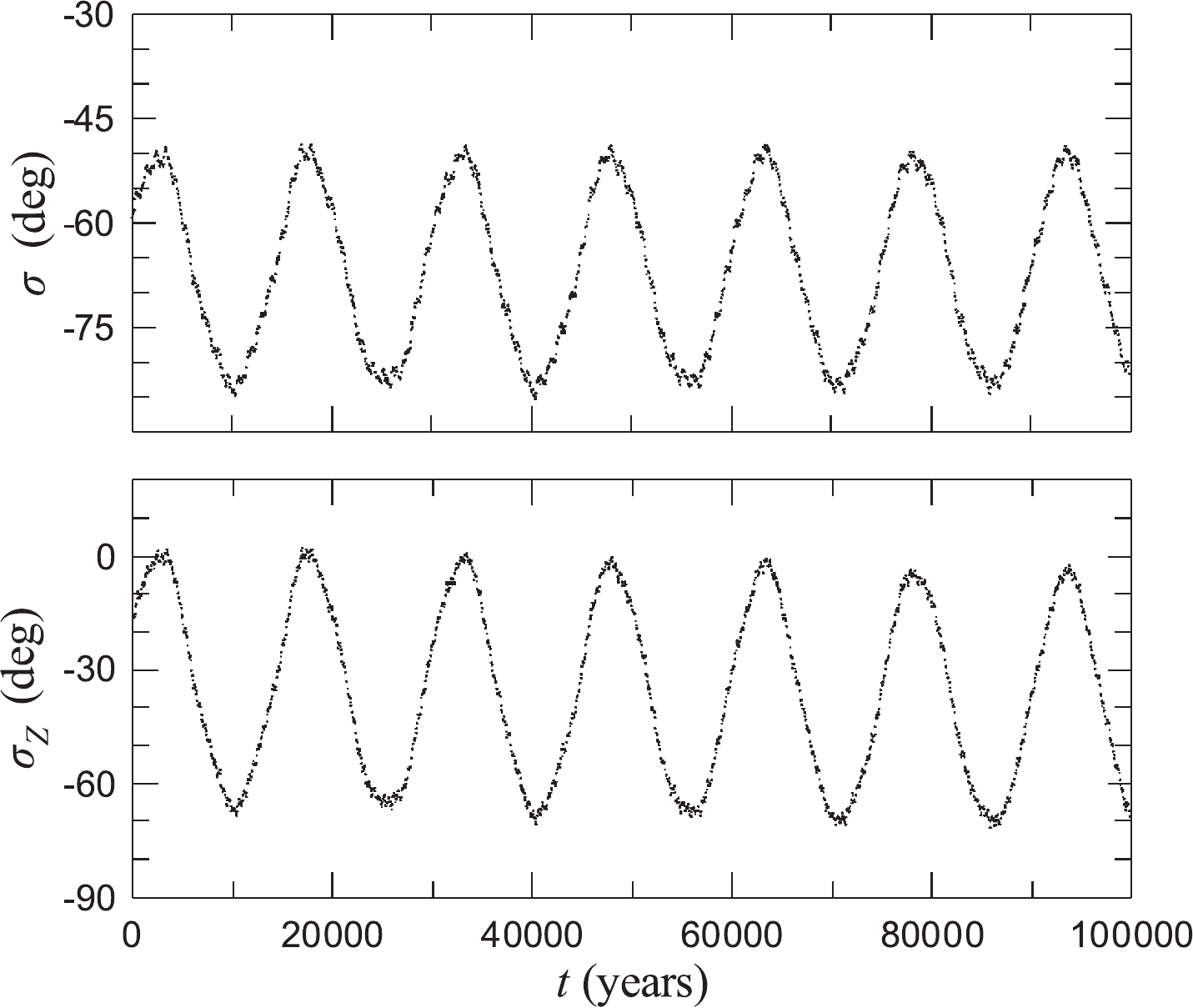} & \quad & \includegraphics[width=0.45\textwidth]{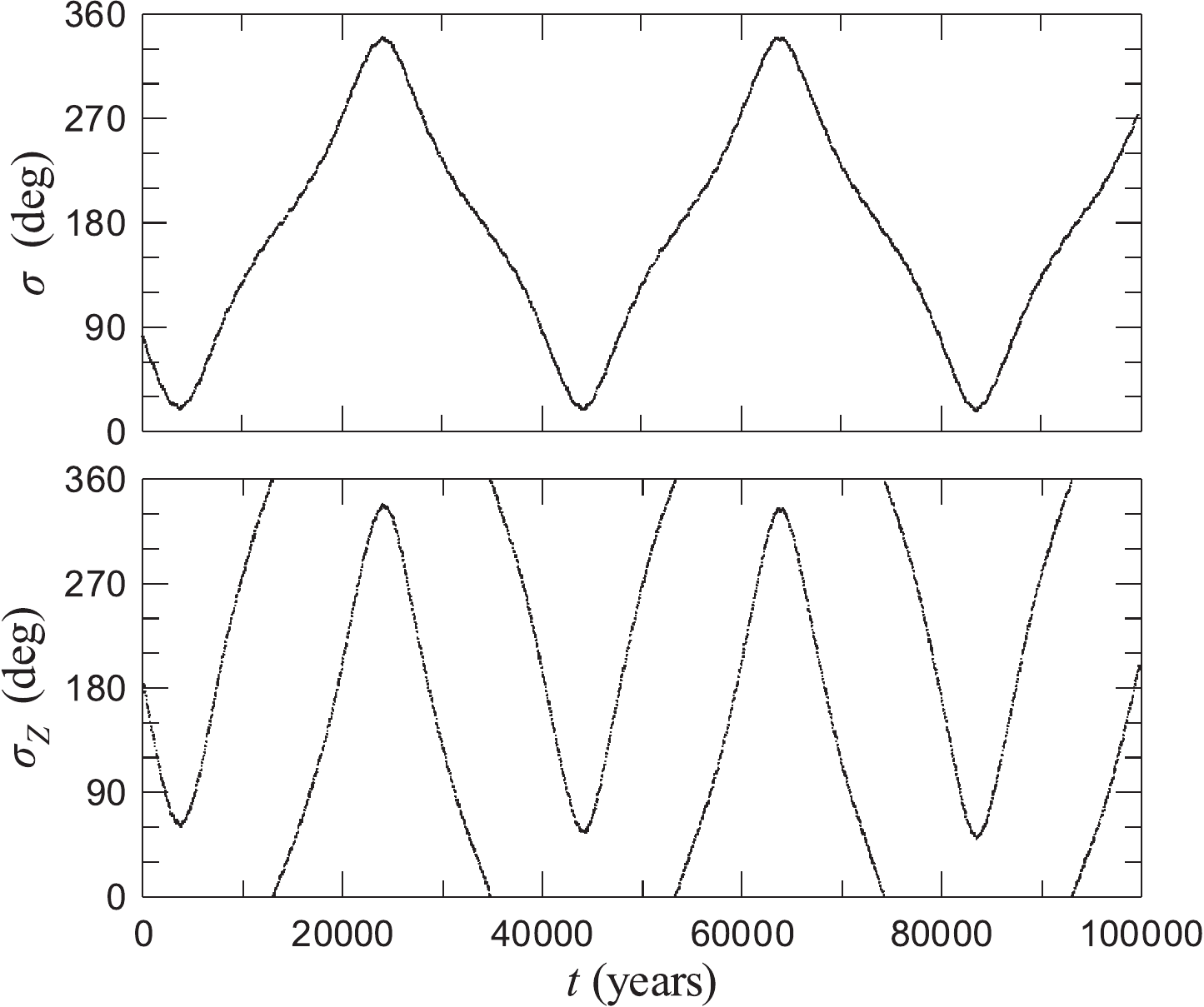} \\
\end{array}$
\caption{Time evolution of $\sigma$~(top) and $\sigma_z$~(bottom) for the TNOs (20161) in the left panel and (130391) in the right panel. These 
correspond to the two different types of $1:2$ TNO dynamics found being described in the text}
\label{tnossz}
\end{figure}

Certainly, the most important manifestation of asymmetric libration in the solar system is the existence of asymmetric, $1:2$-resonant TNOs with Neptune. We integrated all TNOs contained in the latest version of the AstDyS catalogue and found 9 objects in this state. Let us note that, despite the proximity of some TNOs to other 1:$p$ resonances -- out of the very few objects that have been catalogued exterior to the 1:2 MMR, so far --  we did not manage to find any stable librations. However, these nine TNOs show two distinct types of behaviour, similar to those seen in Fig.~\ref{FIGLIBROT}. Six of them show libration of both $\sigma$ and $\sigma_Z$, while the remaining three show a small-amplitude libration of $\sigma$, accompanied by a near-$360^{\circ}$ libration of $\sigma_Z$, which signifies ``horseshoe-type'' motion, around the separatrix that encompasses two stable a.p.o that are mirror images of one another. Two examples are respectively shown in Fig.~\ref{tnossz}.

\begin{figure}
$
\begin{array}{ccc}
\includegraphics[width=0.45\textwidth]{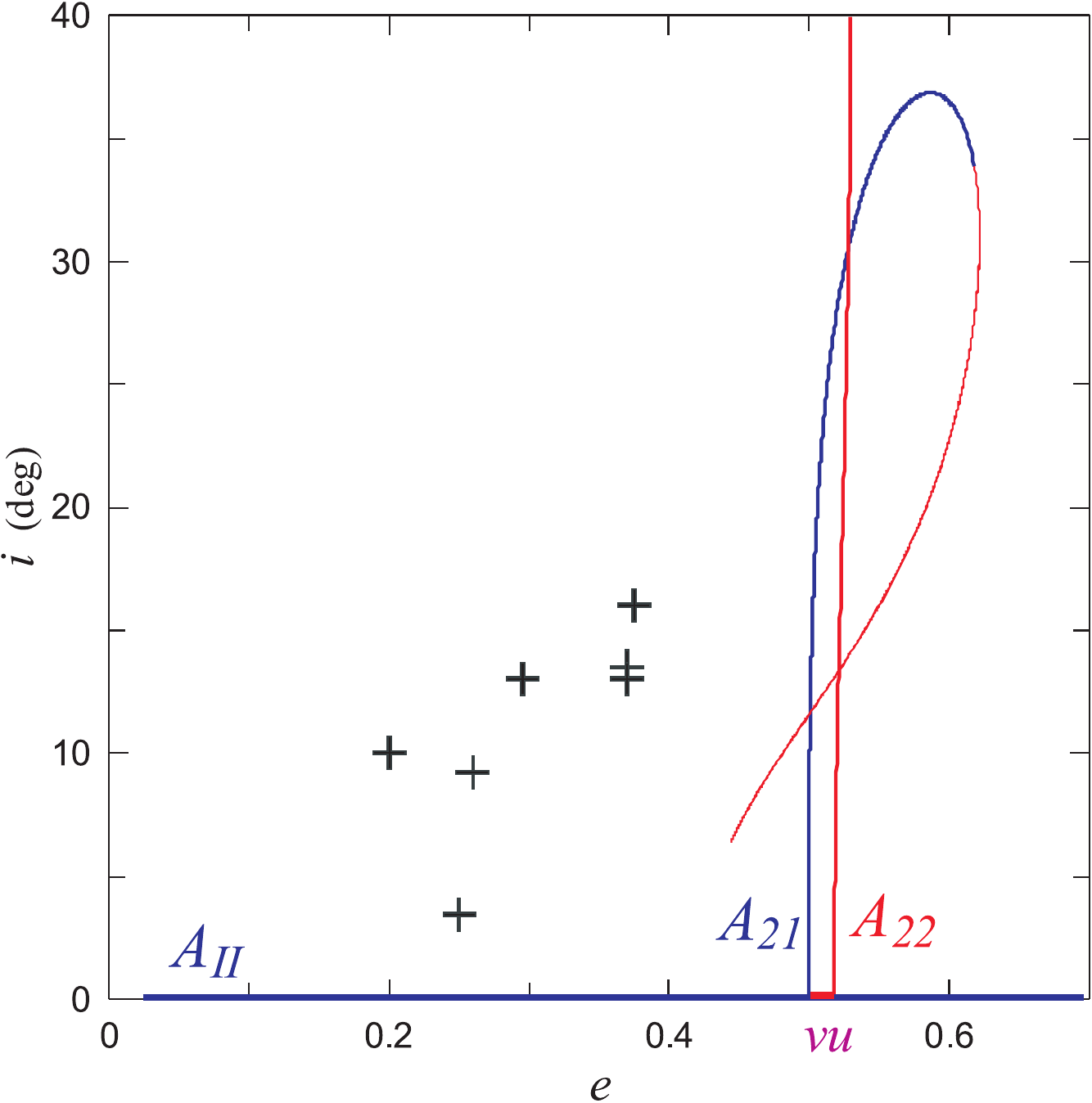} & \quad & \includegraphics[width=0.45\textwidth]{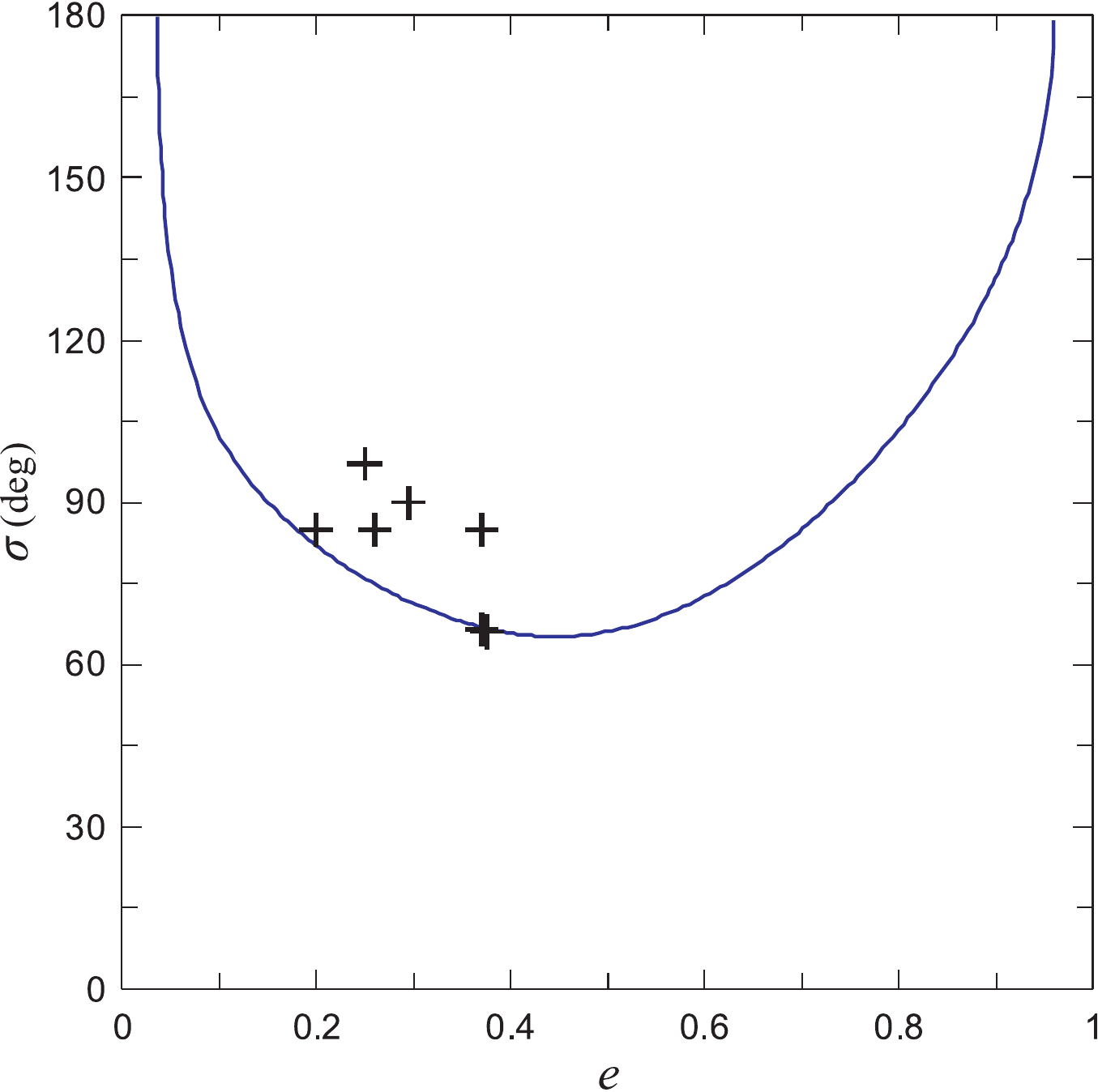} \\
\textnormal{(a)} & \quad & \textnormal{(b)}  
\end{array}$
\caption{ {\bf a} Distribution of TNOs on the $e - i$ plane that showcase $1:2$ resonant asymmetric librations. The asymmetric spatial families are also presented (see also Fig.~\ref{FIGAFAM12})  {\bf b} Distribution of TNOs according to their centre of $\sigma$--libration. The angle $\sigma$ along the planar asymmetric family $A_{II}$ is also shown}
\label{tnosdistr}
\end{figure} 

Note that these asymmetric $1:2$-resonant TNOs are projected on the $(e,i)$ plane relatively far from the corresponding spatial asymmetric families (Fig.~\ref{tnosdistr}). Moreover, their centres of $\sigma$-libration cluster around the values predicted for the asymmetric {\it planar} family in the CRTBP; they are actually on inclined orbits, but  with a moderate $i \leq 18^{\circ}$. This suggests that the dynamics of these TNOs are closer related to the planar, rather than the spatial a.p.o. We will come back to this point in the following section.

\section{Conclusions}

The asymmetric periodic orbits of the planar CRTBP have been studied in many papers in the past and their structure (bifurcation and position in phase space and stability) is well known. It seems that they exist only for $1:p$ ($p=2,3...$) exterior resonances. In this paper, we address the spatial problem and we proceed to computations of inclined asymmetric periodic solutions for the $1:2$, $1:3$, $1:4$ and $1:5$ resonant motion, in the case of the Sun-Neptune-TNO model approximated by the circular restricted three-body problem. Even though the computational approach had been developed by \cite{Markellos78}, only partial results were presented and only for the $1:2$ resonance.

Considering the family $A_{II}$ of planar asymmetric periodic orbits, we computed its vertical stability for each resonance. For symmetric configurations, it is known by \cite{henon73} that vertical critical orbits (v.c.o.) constitute bifurcation points for spatial families of periodic orbits. The computations of \cite{Markellos78} mentioned above were based on the conjecture that asymmetric v.c.o. also consist bifurcation points for spatial asymmetric families.  We showed that along the $1:2$ resonant family $A_{II}$ there exist two v.c.o., located close to each other and assumed as a pair. For the $1:3$, $1:4$ and $1:5$ resonance the planar asymmetric family shows two pairs of v.c.o. 
 
By using a suitable differential correction scheme, we showed that asymmetric v.c.o. are indeed continued from the plane to the three-dimensional space,  for all cases of resonance studied. This seems to verify Markellos' conjecture. However, our numerical results for $1:2$-resonant TNOs suggest that there must exist families of three-dimensional a.p.o. that {\it do not bifurcate from the planar asymmetric v.c.o.}. This is indicated by the fact that TNOs show libration in both $\sigma$ and $\sigma_z$, but lie on a different part of the $(e,i)$ diagrams. This subject certainly merits additional investigation. 

Our continuation procedure provided families of spatial asymmetric periodic orbits, generally, up to large inclination values. Here, we presented results only for prograde orbits ($i<90^\circ$). Some of these families terminate to families of spatial symmetric periodic orbits. Also, the centres of libration of the resonant angles  $\sigma$ and $\sigma_Z$ can vary either marginally or significantly along the asymmetric families.  

Our computations of the linear stability, although performed with high accuracy, they are not always conclusive, possibly due to the small value of the mass parameter used and due to the asymmetry. This is why we resorted to computing the long-term evolution of deviation vectors. We found that at each pair of v.c.o. and for all resonances studied, one family starts with stable orbits and one with unstable ones. The families that start as unstable remain unstable until the termination of the family, or along the whole range of values of the continuation parameter. The family that starts as stable, continues having stable orbits up to approximately $30^\circ$. Then, its orbits become unstable. However, instability of asymmetric orbits is ``weak'', in the sense that it is not associated with strongly chaotic orbits that leave the vicinity of the unstable a.p.o. quickly. For the unstable periodic orbits there is always a ``separatrix'' solution and, for orbits close to this solution, the resonant angle $\sigma$ librates, but $\sigma_Z$ either librates with amplitude $\sim 360^\circ$ or rotates; we show that this behaviour is found among real TNOs, in a model that accounts for the gravity of all major planets. \\

\noindent
{\bf Acknowledgements.} \\
The work of KIA was supported by the Fonds de la Recherche Scientifique-FNRS under Grant No. T.0029.13 (``ExtraOrDynHa'' research project) and the University of Namur.

\noindent
{\bf Conflict of Interest.} The work of K.I. Antoniadou was supported by the Fonds de la Recherche Scientifique-FNRS under Grant No. T.0029.13 ("ExtraOrDynHa" research project) and the University of Namur. The authors G.Voyatzis and K.Tsiganis declare that they have no conflict of interest.

\bibliographystyle{plainnat}
\bibliography{gvnbib} 

\begin{thebibliography}{29}
\providecommand{\natexlab}[1]{#1}
\providecommand{\url}[1]{\texttt{#1}}
\expandafter\ifx\csname urlstyle\endcsname\relax
  \providecommand{\doi}[1]{doi: #1}\else
  \providecommand{\doi}{doi: \begingroup \urlstyle{rm}\Url}\fi

\bibitem[Antoniadou and Voyatzis(2013)]{av13b}
K.~I. Antoniadou and G.~Voyatzis.
\newblock 2/1 resonant periodic orbits in three dimensional planetary systems.
\newblock \emph{Celestial Mechanics and Dynamical Astronomy}, 115:\penalty0
  161--184, 2013.

\bibitem[{Antoniadou} and {Voyatzis}(2014)]{av14b}
K.~I. {Antoniadou} and G.~{Voyatzis}.
\newblock {Resonant periodic orbits in the exoplanetary systems}.
\newblock \emph{Astrophysics and Space Science}, 349:\penalty0 657--676, 2014.

\bibitem[Antoniadou et~al.(2011)Antoniadou, Voyatzis, and Kotoulas]{avk11}
K.~I. Antoniadou, G.~Voyatzis, and T.~Kotoulas.
\newblock On the bifurcation and continuation of periodic orbits in the three
  body problem.
\newblock \emph{International Journal of Bifurcation and Chaos}, 21:\penalty0
  2211, 2011.

\bibitem[Beaug{\'e}(1994)]{Beauge94}
C.~Beaug{\'e}.
\newblock Asymmetric librations in exterior resonances.
\newblock \emph{Celestial Mechanics and Dynamical Astronomy}, 60:\penalty0
  225--248, 1994.

\bibitem[{Beauge} and {Ferraz-Mello}(1994)]{bf94}
C.~{Beauge} and S.~{Ferraz-Mello}.
\newblock Capture in exterior mean-motion resonances due to poynting-robertson
  drag.
\newblock \emph{Icarus}, 110:\penalty0 239--260, 1994.
\newblock \doi{10.1006/icar.1994.1119}.

\bibitem[{Berry}(1978)]{Berry1978}
M.~V. {Berry}.
\newblock {Regular and irregular motion}.
\newblock In \emph{Topics in nonlinear dynamics: A tribute to Sir Edward
  Bullard, American Institute of Physics}, pages 16--120, 1978.

\bibitem[Broucke(1969)]{Broucke69}
R.~Broucke.
\newblock Stability of periodic orbits in the elliptic, restricted three-body
  problem.
\newblock \emph{AIAA Journal}, 7:\penalty0 1003--1009, 1969.

\bibitem[Bruno(1994)]{Bruno94}
A.~D. Bruno.
\newblock \emph{The Restricted 3-Body Problem: Plane Periodic Orbits}.
\newblock Walter de Gruyter, Berlin, New York, 1994.

\bibitem[Chambers(1999)]{Chambers99}
J.E. Chambers.
\newblock A hybrid symplectic integrator that permits close encounters between
  massive bodies.
\newblock \emph{Monthly Notices of the Royal Astronomical Society},
  304:\penalty0 793--799, 1999.

\bibitem[{Giuppone} et~al.(2010){Giuppone}, {Beaug{\'e}}, {Michtchenko}, and
  {Ferraz-Mello}]{Giuppone10}
C.~A. {Giuppone}, C.~{Beaug{\'e}}, T.~A. {Michtchenko}, and S.~{Ferraz-Mello}.
\newblock {Dynamics of two planets in co-orbital motion}.
\newblock \emph{Monthly Notices of the Royal Astronomical Society},
  407:\penalty0 390--398, 2010.
\newblock \doi{10.1111/j.1365-2966.2010.16904.x}.

\bibitem[Hadjidemetriou(1993)]{Hadjidem93}
J.~D. Hadjidemetriou.
\newblock Resonant motion in the restricted three body problem.
\newblock \emph{Celestial Mechanics and Dynamical Astronomy}, 56:\penalty0
  201--219, 1993.

\bibitem[Hadjidemetriou and Voyatzis(2011)]{hv11}
J.~D. Hadjidemetriou and G.~Voyatzis.
\newblock Different types of attractors in the three body problem perturbed by
  dissipative terms.
\newblock \emph{International Journal of Bifurcation and Chaos}, 21:\penalty0
  2195--2209, 2011.

\bibitem[H\'enon(1973)]{henon73}
M.~H\'enon.
\newblock Vertical stability of periodic orbits in the restricted problem. i.
  equal masses.
\newblock \emph{Astronomy and Astrophysics}, 28:\penalty0 415, 1973.

\bibitem[{H\'enon}(1997)]{henon97}
M.~{H\'enon}.
\newblock \emph{{Generating Families in the Restricted Three-Body Problem}}.
\newblock Springer-Verlag, 1997.

\bibitem[{Ichtiaroglou} et~al.(1989){Ichtiaroglou}, {Katopodis}, and
  {Michalodimitrakis}]{ikm89}
S.~{Ichtiaroglou}, K.~{Katopodis}, and M.~{Michalodimitrakis}.
\newblock Periodic orbits in the three-dimensional planetary systems.
\newblock \emph{Journal of Asttrophysics and Astronomy}, 10:\penalty0 367--380,
  1989.

\bibitem[{Jewitt}(1999)]{Jewitt99}
D.~{Jewitt}.
\newblock {Kuiper Belt Objects}.
\newblock \emph{Annual Review of Earth and Planetary Sciences}, 27:\penalty0
  287--312, 1999.
\newblock \doi{10.1146/annurev.earth.27.1.287}.

\bibitem[Kotoulas and Voyatzis(2005)]{kv05}
T.~A. Kotoulas and G.~Voyatzis.
\newblock Three dimensional periodic orbits in exterior mean motion resonances
  with neptune.
\newblock \emph{Astronomy and Astrophysics}, 441:\penalty0 807--814, 2005.

\bibitem[{Lykawka} and {Mukai}(2007)]{LykawkaMukai07}
P.~S. {Lykawka} and T.~{Mukai}.
\newblock Dynamical classification of trans-neptunian objects: Probing their
  origin, evolution, and interrelation.
\newblock \emph{Icarus}, 189:\penalty0 213--232, 2007.
\newblock \doi{10.1016/j.icarus.2007.01.001}.

\bibitem[{Markellos}(1977)]{Markellos77}
V.~V. {Markellos}.
\newblock Bifurcations of plane with three-dimensional asymmetric periodic
  orbits in the restricted three-body problem.
\newblock \emph{Monthly Notices of the Royal Astronomical Society},
  180:\penalty0 103--116, 1977.
\newblock \doi{10.1093/mnras/180.2.103}.

\bibitem[{Markellos}(1978)]{Markellos78}
V.~V. {Markellos}.
\newblock Asymmetric periodic orbits in three dimensions.
\newblock \emph{Monthly Notices of the Royal Astronomical Society},
  184:\penalty0 273--281, August 1978.
\newblock \doi{10.1093/mnras/184.2.273}.

\bibitem[{Message}(1958)]{Message58}
P.~J. {Message}.
\newblock {The search for asymmetric periodic orbits in the restricted problem
  of three bodies}.
\newblock \emph{The Astronomical Journal}, 63:\penalty0 443, 1958.
\newblock \doi{10.1086/107804}.

\bibitem[{Murray} and {Dermott}(1999)]{murraydermott}
C.~D. {Murray} and S.~F. {Dermott}.
\newblock \emph{{Solar system dynamics}}.
\newblock Cambridge University Press, 1999.

\bibitem[{Szebehely}(1967)]{sze}
V.~{Szebehely}.
\newblock \emph{{Theory of orbits. The restricted problem of three bodies}}.
\newblock Academic Press, 1967.

\bibitem[Taylor(1983{\natexlab{a}})]{Taylor83a}
D.~B. Taylor.
\newblock Families of asymmetric periodic solutions of the restricted problem
  of three bodies for the sun-jupiter mass ratio and their relationship with
  the symmetric families.
\newblock \emph{Celestial Mechanics}, 29:\penalty0 51--74, 1983{\natexlab{a}}.

\bibitem[Taylor(1983{\natexlab{b}})]{Taylor83b}
D.~B. Taylor.
\newblock Evolution with the mass parameter of families of asymmetric periodic
  solutions of the restricted three body problem.
\newblock \emph{Celestial Mechanics}, 29:\penalty0 75--98, 1983{\natexlab{b}}.

\bibitem[Voyatzis and Kotoulas(2005)]{vk05}
G.~Voyatzis and T.~Kotoulas.
\newblock Planar periodic orbits in exterior resonances with neptune.
\newblock \emph{Planetary and Space Science}, 53:\penalty0 1189--1199, 2005.

\bibitem[Voyatzis et~al.(2005)Voyatzis, Kotoulas, and Hadjidemetriou]{vkh05}
G.~Voyatzis, T.~Kotoulas, and J.~D. Hadjidemetriou.
\newblock Symmetric and nonsymmetric periodic orbits in the exterior mean
  motion resonances with neptune.
\newblock \emph{Celestial Mechanics and Dynamical Astronomy}, 91:\penalty0
  191--202, 2005.

\bibitem[{Winter} and {Murray}(1997)]{winter97}
O.~C. {Winter} and C.~D. {Murray}.
\newblock {Resonance and Chaos. II. Exterior resonances and asymmetric
  libration}.
\newblock \emph{Astronomy and Astrophysics}, 328:\penalty0 399--408, 1997.

\bibitem[{Zagouras} et~al.(1996){Zagouras}, {Perdios}, and {Ragos}]{Zagouras96}
C.~G. {Zagouras}, E.~{Perdios}, and O.~{Ragos}.
\newblock {New kinds of asymmetric periodic orbits in the restricted three-body
  problem}.
\newblock \emph{Astrophysics and Space Science}, 240:\penalty0 273--293, 1996.
\newblock \doi{10.1007/BF00639592}.

\end{thebibliography}
\end{document}